\documentclass[sigconf]{acmart}

\usepackage{booktabs} % For formal tables
\usepackage{multirow}
\usepackage{subfigure}
\usepackage{verbatim}
\usepackage{CJKutf8}

\usepackage{makecell}
\usepackage[utf8]{inputenc}
\setcopyright{rightsretained}
\usepackage{flushend}
\acmDOI{10.475/123_4}

\acmISBN{123-4567-24-567/08/06}

\acmConference[WWW 2021]{The Web Conference}{April 2021}{Ljubljana, Slovenia}
\acmYear{2021}
\copyrightyear{2021}
\acmArticle{4}
\acmPrice{15.00}

\editor{Jennifer B. Sartor}
\editor{Theo D'Hondt}

\begin{document}

\title{FedCTR: Federated Native Ad CTR Prediction with Multi-Platform User Behavior Data}
\fancyhead{}
%\subtitle{FedCTR}
%\subtitlenote{The full version of the author's guide is available as
%  \texttt{acmart.pdf} document}

\author{Chuhan Wu$^1$, Fangzhao Wu$^2$, Tao Di$^3$, Yongfeng Huang$^1$, Xing Xie$^2$}

\affiliation{%
  \institution{$^1$Tsinghua University, Beijing, China \\ $^2$Microsoft Research Asia, Beijing, China\\ $^3$Microsoft, Seattle, Washington, United States}
} 
\email{{wuchuhan15,wufangzhao}@gmail.com, Tao.Di@microsoft.com, yfhuang@.tsinghua.edu.cn, xing.xie@microsoft.com}

\begin{abstract}

Native ad is a popular type of online advertisement which has similar forms with the native content displayed on websites.
Native ad CTR prediction is useful for improving user experience and platform revenue.
However, it is a challenging task due to the lack of explicit user intent, and  users' behaviors on the platform with native ads may not be sufficient to infer their interest in ads.
Fortunately, user behaviors exist on many online platforms and they can provide complementary information for user interest mining.
Thus, leveraging multi-platform user behaviors is useful for native ad CTR prediction.
However, user behaviors are highly privacy-sensitive and the behavior data on different platforms cannot be directly aggregated due to user privacy concerns and data protection regulations like GDPR.
Existing CTR prediction methods usually require centralized storage of user behavior data for user modeling and cannot be directly applied to the CTR prediction task with multi-platform user behaviors.
In this paper, we propose a federated native ad CTR prediction method named FedCTR, which can learn user interest representations from their behaviors on multiple platforms in a privacy-preserving way without the need of centralized storage.
On each platform a local user model is used to learn user embeddings from the local user behaviors on that platform.
The local user embeddings from different platforms are uploaded to a server for aggregation, and the aggregated user embeddings are sent to the ad platform for CTR prediction.
Besides, we apply local differential privacy (LDP) and differential privacy (DP) techniques to the local and aggregated user embeddings respectively for better privacy protection.
Moreover, we propose a federated framework for collaborative model training with distributed models and distributed user behaviors.
Extensive experiments on real-world dataset demonstrate that the proposed method can effectively leverage multi-platform user behaviors for native ad CTR prediction in a privacy-preserving manner.

\end{abstract}



\maketitle

\section{Introduction}

\begin{figure}[!t]
  \centering
    \includegraphics[width=0.98\linewidth]{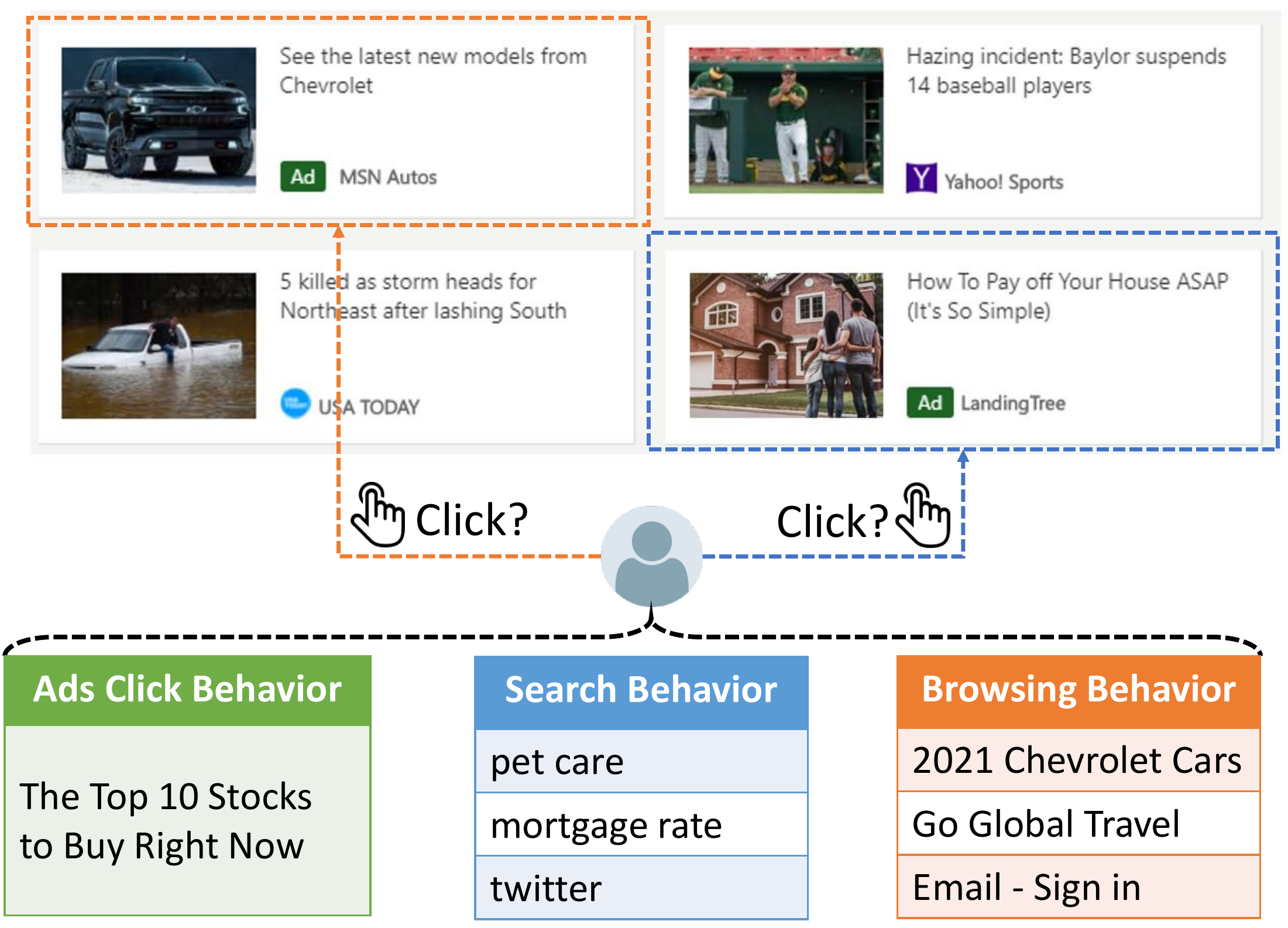}
 \caption{An illustrative example of several ads and user behaviors on different platforms.}

  \label{fig.exp}
\end{figure}

Native ad is a popular form of online advertisements that has similar style and function with the native content displayed on online platforms such as news and video websites~\cite{matteo2015native}.
An illustrative example of native ads on the homepage of MSN News\footnote{https://www.msn.com/en-us} is shown in Fig.~\ref{fig.exp}.
We can see that except for a sign of ``Ad'', the appearance of the embedded native ads is very similar with the listed news articles.
Due to the reduction of ad recognition, native ads can better attract users' attentions, and have gained increasing popularity in many online platforms~\cite{wojdynski2016going}.
Therefore, accurate click-through rate (CTR) prediction of native ads is an important task for online advertising, which can help improve users' experience by recommending ads that they are interested in as well as improve the revenue of online websites by attracting more ad clicks~\cite{chen2016deep}.

Although the CTR prediction for search ads and display ads has been widely studied~\cite{li2019graph,zhou2019deep}, the research on native ads is very limited~\cite{an2019neural}.
Compared with search ads which are distributed based on inferring users' intents from their search queries, there is no explicit user intent for native ads, making it more difficult to predict the click probability.
In addition, the CTR prediction of display ads which are usually presented on e-commerce platforms has the advantage of targeting users' interest in products based on their various behavior records such as browsing, preferring and purchasing.
However, users' behaviors on the platform where native ads are displayed  usually cannot provide sufficient information for inferring their interest in ads, such as the news reading behaviors at online news websites.

Fortunately, users' online behaviors exist in many different platforms, and they can provide various clues to infer user interest in different aspects.
For example, in Fig.~\ref{fig.exp}, the search behaviors on search engine platform (e.g., ``mortgage rate'') indicate that this user may have interest in buying a new house or applying for housing loan, and she may click the native ad with title ``How to Pay Off Your House ASAP''.
In addition, from her webpage browsing behaviors we can infer that this user may be interested in cars since she browsed a webpage with title ``2021 Chevrolet Cars'', and it is appropriate to display the native ad ``See the latest new models from Chevrolet'' to her.
Thus, incorporating user behaviors on multiple platforms is useful for modeling user interest more accurately and can benefit native ad CTR prediction, which has been validated by existing studies~\shortcite{an2019neural}.
For example, An et al.~\shortcite{an2019neural} found that combining users' searching behaviors and webpage browsing behaviors can achieve better performance on native ad CTR prediction than using single kind of user behaviors.
However, online user behaviors such as the queries they searched and the webpages they browsed are highly privacy-sensitive.
Thus, the behavior data on different platforms cannot be directly aggregated into a single server or exchanged among different platforms due to users' privacy concerns and the user data protection regulations like GDPR\footnote{https://gdpr-info.eu}. 
Most existing CTR prediction methods rely on centralized storage of user behavior data, making them difficult to be applied to the native ad CTR prediction task with multi-platform user behaviors.

In this paper, we propose a native ad CTR prediction method named FedCTR, which is based on federated learning and can incorporate the user behavior data on different platforms to model user interest in a privacy-preserving way without the need of centralized storage.
Specifically, we learn unified user representations from different platforms in a federated way.
On each platform that participates in user representation learning, a local user model is used to learn the local user embeddings from the user behavior logs on that platform.
The learning of user embeddings on different platforms is coordinated by a user server, and these local embeddings are uploaded to this server for aggregation.
The aggregated unified user embeddings are further sent to the ad platform for CTR prediction.
Thus, the FedCTR method can exploit the user information on different platforms for native ad CTR prediction.
Since the raw user behavior logs never leave the local platform and only user embeddings are uploaded, user privacy can be protected to a certain extent in the FedCTR method. 
In addition, we apply local differential privacy (LDP)~\cite{ren2018textsf} and differential privacy (DP)~\cite{dwork2008differential} techniques to the local and aggregated user embeddings respectively before sending them, which can achieve better privacy protection at a small cost of CTR prediction performance.
Since in FedCTR the user behavior data and model parameters are located on different platforms, it is a non-trivial task to train the model of FedCTR without exchanging the privacy-sensitive user behavior data across different platforms.
We propose a privacy-preserving framework based on federated learning to address this issue. 
In our framework, only user embeddings and model gradients are communicated among different platforms, thus the risk of user privacy leakage can be effectively reduced at the model training stage.
Extensive experiments are conducted on a real-world dataset collected from the user logs on the native ads in MSN News website\footnote{https://www.msn.com/en-us}.
The results validate that our approach can improve the performance of native ad CTR prediction via exploiting the multi-platform user behaviors and meanwhile effectively protect user privacy.

The main contributions of this work include:
\begin{itemize}
    \item We propose a privacy-preserving native ad CTR prediction
method FedCTR which can exploit multi-platform user behaviors for user interest modeling without centralized storage.
    
    \item We propose a federated model training framework to train models in FedCTR where user behavior data and model parameters are distributed on different platforms.
    
    \item We conduct extensive experiments on  real-world dataset to verify the effectiveness of the proposed method in both CTR prediction and privacy protection. 
\end{itemize}

\section{Related Work}\label{sec:RelatedWork}

\subsection{CTR Prediction}

Native ad is a special kind of display ads which has similar form with the native content displayed in online websites~\cite{matteo2015native}.
CTR prediction for display ads has been extensively studied~\cite{richardson2007predicting,cheng2016wide,guo2017deepfm,zhou2018deep,zhou2019deep}.
Different from search ads where the search query triggering the impression of the ads can provide clear user intent, in display ads there is no explicit intent from the users~\cite{zhou2018deep}.
Thus, it is very important for display ad CTR prediction to model user interest from users' historical behaviors.
Many existing CTR prediction methods rely on handcrafted features to represent users and ads, and they focus on capturing the interactions between these features to estimate the relevance between users and ads~\cite{richardson2007predicting,chakrabarti2008contextual,juan2016field,cheng2016wide,guo2017deepfm,zhou2018deep,pan2018field,lian2018xdeepfm}.
For example, Cheng et al.~\shortcite{cheng2016wide} proposed a Wide\&Deep model that integrates a wide linear channel with cross-product and a deep neural network channel to capture feature interactions for CTR prediction.
They represented users' interest with their demographic features, device features and the features extracted from their historical impressions.
Guo et al.~\shortcite{guo2017deepfm} proposed a DeepFM model that uses a combination of factorization machines and deep neural networks to model feature interactions.
They used ID and category features to represent ads, and used the feature collections of historical clicked items to represent users.
However, the design of handcrafted features used in these methods usually requires massive domain knowledge, and handcrafted features may not be optimal in modeling user interest.
There are also several methods for display ads CTR prediction that use deep learning techniques to learn user interest representations from their behaviors on e-commerce platforms~\cite{pi2019practice,feng2019deep,li2020interpretable}.
For example, Zhou et al.~\shortcite{zhou2018deep} proposed a deep interest network (DIN) that learns representations of users from the  items they  have interacted with on the e-commerce platform based on the relatedness between those items and the candidate ads.
In~\cite{zhou2019deep}, an improved version of DIN named deep interest evolution network (DIEN) was proposed, which models users from their historical behaviors on items via a GRU network with attentional update gate.
These deep learning based CTR prediction methods for display ads on e-commerce platform usually rely on the user behaviors on the same platform to model user interest.
However, besides the e-commerce platform, native ads are widely displayed on many other online platforms such as news websites~\cite{an2019neural}.
The user behaviors on these platforms may have insufficient clues to infer user interest in ads.
Thus, these CTR prediction methods designed for display ads on e-commerce platform may be not optimal for native ads.

There are only a few studies for native ad CTR prediction~\cite{parsana2018improving,an2019neural}.
For example, An et al.~\shortcite{an2019neural} proposed to model user interest for native ad CTR prediction from their search queries and browsed webpages.
These studies found that incorporating multi-platform user behaviors can model user interest more accurately than using single-platform user behaviors.
These methods usually require centralized storage of multi-platform user behaviors.
However, user behavior data is highly privacy-sensitive, and cannot be directly aggregated across platforms due to privacy concerns and user data protection regulations like GDPR.
Different from these methods, our proposed FedCTR method can exploit user behaviors on different platforms for native ad CTR prediction via federated learning, which can remove the need to centralized storage of user behavior data and achieve better user privacy protection.

\subsection{Federated Learning}

The learning of user representation in our proposed FedCTR method with multi-platform user behavior data is based on federated learning~\cite{mcmahan2017communication}.
Federated learning is a recently proposed machine learning technique, which can learn a shared model from the private data of massive users in a privacy-preserving way~\cite{hardy2017private,nock2018entity,liu2019communication,yang2019federated,feng2020multi}.
Instead of directly uploading the private user data to a central server for model training, in federated learning the user data is locally stored on different user devices such as smartphones and personal computers.
Each user device has a copy of the local model and computes the model updates based on local user interactions. 
The model updates from a large number of users are uploaded to the server and aggregated into a single one for global model update~\cite{mcmahan2017communication}.
Then the new global model is delivered to user devices for local model update, and this process iterates for multiple rounds.
Since the model updates contain much less information than the raw user data, federated learning can provide an effective way to exploit the private data of different users and protect their privacy at the same time~\cite{mcmahan2017communication}.
Based on the idea of federated learning, Jiang et al.~\shortcite{jiang2019federated} proposed a federated topic modeling approach to train topic models from the corpus owned by different parties.
In these federated learning methods, the samples for model training are distributed on different clients, and each client shares the same feature space.
Different from these methods, in the task of native ad CTR prediction with multi-platform user behaviors, the user behavior data of each sample is distributed on different platforms.
These platforms may contain the same sample but they can only see part of the user feature space.
Thus, the problem studied in this work is quite different from the existing federated learning methods.
To our best knowledge, this is the first work which applies federated learning to privacy-preserving native ad CTR prediction with multi-platform user behaviors.

\section{Methodology}\label{sec:Model}

In this section, we first introduce the details of our federated native ad CTR prediction method (FedCTR).
Then, we introduce the framework to train the FedCTR model where  behavior data and model parameters are distributed on different platforms.

\begin{figure*}[!t]
  \centering
    \includegraphics[width=0.85\linewidth]{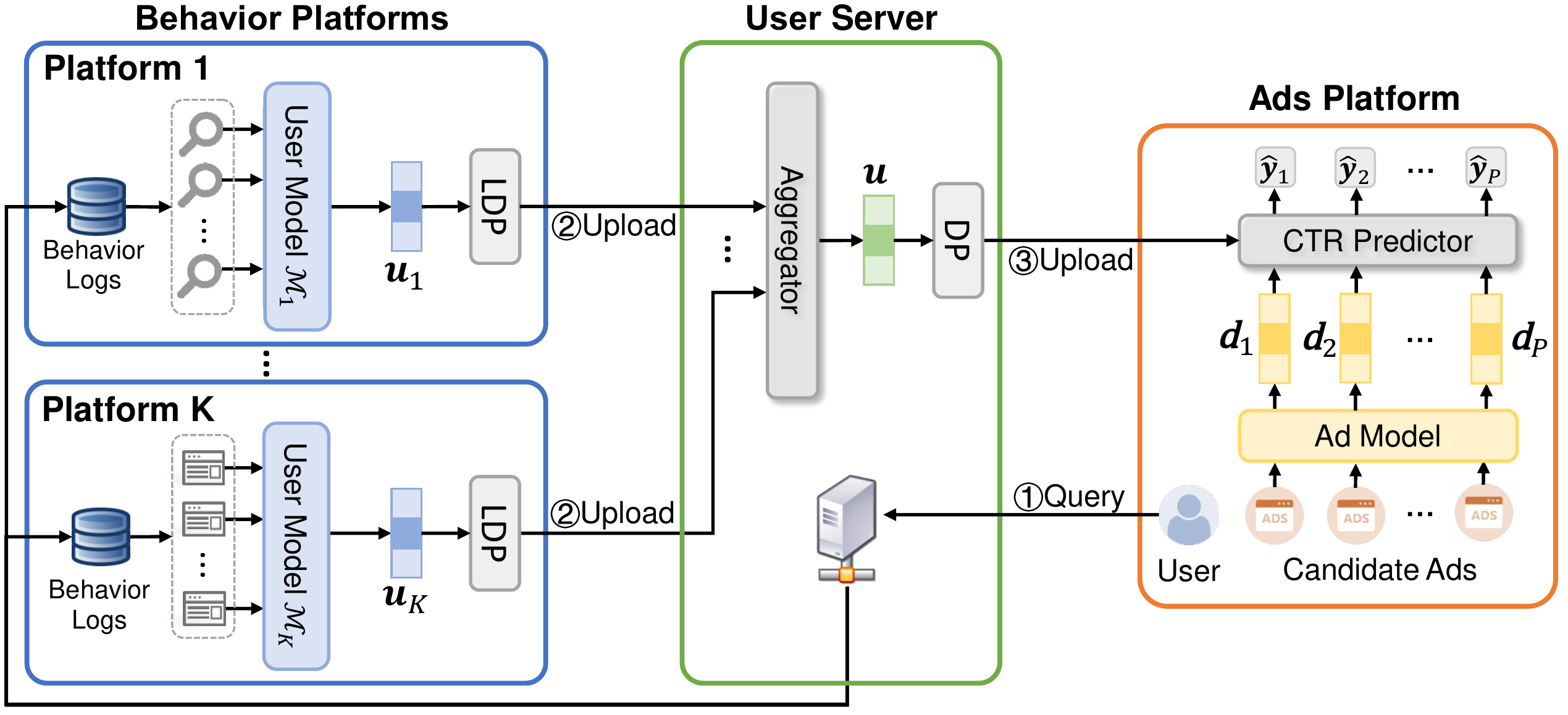}
  \caption{The architecture of the \textit{FedCTR} method.}
  \label{fig.model}
\end{figure*}

\subsection{FedCTR for Native Ad CTR Prediction}

\subsubsection{Overall Framework}\label{section.framework}

The architecture of the proposed FedCTR method is shown in Fig.~\ref{fig.model}.
In FedCTR, user behaviors on multiple online platforms are used to infer user interest for native ad CTR prediction, and the behavior data cannot be directly uploaded to a server or exchanged across different platforms due to privacy concerns.
Concretely, there are three major modules in the FedCTR framework.
The first module is ad platform, which is used to predict the CTR scores of a set of native ads using a CTR predictor.
It computes the probability score of a target user $u$ clicking a candidate ad $d$ based on their representations $\mathbf{u}$  and $\mathbf{d}$, which is formulated as  $\hat{y}=f_{CTR}(\mathbf{u},\mathbf{d}; \Theta_C)$, where $\Theta_C$ represents the parameters of the CTR predictor.
The representation of the ad $d$ is computed by an \textit{ad model} based on its ID and text, which is formulated as $\mathbf{d}=f_{ad}(ID, text; \Theta_D)$, where $\Theta_D$ denotes the model parameters of the \textit{ad model}. 
The second module consists of $K$ behavior platforms.
Each platform has a \textit{user model} to learn local user representations based on its stored local user behaviors, such as search queries on the search engine platform and browsed webpages on the web browsing platform.
For the $i$-th behavior platform, the learning of local user representation is formulated as $\mathbf{u}_i=f_{user}^i(behaviors; \Theta_{U_i})$, where $\Theta_{U_i}$ denotes the parameter of the \textit{user model} maintained by this platform. 
The third module is a user server, which is responsible for coordinating multiple behavior platforms to learn local user embeddings according to the query of the  ad platform and aggregating them into a unified user representation $\mathbf{u}$, which is formulated as $\mathbf{u}=f_{agg}(\mathbf{u}_1, \mathbf{u}_2, ..., \mathbf{u}_K; \Theta_A)$, where $\Theta_A$ is the aggregator model parameters.
The aggregated user embeddings are further sent to the ad platform.
Next, we introduce each module in detail.

In the  ad platform, assume there is a set of candidate ads, denoted as $\mathcal{D}=[d_1, d_2, ..., d_P]$.
Each ad has an ID, a title and a description.
There is an \textit{ad model} in the ad platform that learns representations of ads from their ID, title and description.
When a user $u$ visits the website where native ads are displayed, the  ad platform is called to compute the personalized CTR scores of the candidate ads for this user.
It sends the ID of this user to the user server to query her embeddings inferred from her behaviors on multiple platforms which encode her personalized interest information.
When the  ad platform receives the user embedding from the user server, it uses a \textit{CTR predictor} to compute the ranking scores of the candidate ads based on user embeddings $\mathbf{u}$ and embeddings of candidate ads $[\mathbf{d}_1, \mathbf{d}_2, ..., \mathbf{d}_P]$  using $f_{CTR}(\cdot)$, which are denoted as $[\hat{y}_1, \hat{y}_2, ..., \hat{y}_P]$.

The user server is responsible for user embedding generation by coordinating multiple user behavior platforms.
When it receives a  user embedding query from the  ad platform, it will use the user ID to query the $K$ behavior platforms to learn local user embeddings based on the local user behavior logs.
After it receiving the local user embeddings $[\mathbf{u}_1, \mathbf{u}_2, ..., \mathbf{u}_K]$ from different behavior platforms, it uses an \textit{Aggregator} model with the function $f_{agg}(\cdot)$ to aggregate the $K$ local user embeddings into a unified one $\mathbf{u}$, which takes the relative importance of different kinds of behaviors into consideration.
Since the unified user embedding $\mathbf{u}$ may still contain some private information of user behaviors, in order to better protect user privacy we apply differential privacy (DP) technique~\cite{dwork2008differential} to $\mathbf{u}$ by adding Laplacian noise with strength $\lambda_{DP}$ to $\mathbf{u}$.
Then the user server sends the perturbed user embedding $\mathbf{u}$ to the  ad platform for personalized CTR prediction.

The behavior platforms are responsible for learning user embeddings from their local behaviors.
When a behavior platform receives the user embedding query of user $u$, it will retrieve the behaviors of this user on this platform (e.g., search queries posted to the search engine platform), which are denoted as $[d_1, d_2, ..., d_M]$, where $M$ is the number of behaviors.
Then, it uses a neural \textit{user model} to learn the local user embedding $\mathbf{u}_i$ from these behaviors.
The user embedding $\mathbf{u}_i$ can capture the user interest information encoded in user behaviors.
Since the local user embedding may also contain some private information of the user behaviors on the $i$-th behavior platform, we apply local differential privacy (LDP)~\cite{ren2018textsf} by adding Laplacian noise with strength $\lambda_{LDP}$ to each local user embedding so as to better protect user privacy. 
Then, the behavior platform uploads the perturbed local user embedding $\mathbf{u}_i$ to the user server for aggregation.

Next, we provide some discussions on the privacy protection of the proposed FedCTR method.
First, in FedCTR the raw user behavior data never leaves the behavior platforms where it is stored, and only the user embeddings learned from multiple behaviors using neural user models are uploaded to user server.
According to the data processing inequality~\cite{mcmahan2017communication}, the private information conveyed by these local user embeddings is usually much less than the raw user behaviors.
Thus, the user privacy can be effectively protected.
Second, the user server aggregates the local user embeddings from different platforms into a unified one and sends it to the ad platform. 
It is very difficult for the ad platform to infer a specific user behavior on a specific platform from this aggregated user embedding.
Third, we apply the local differential privacy technique to the local user embeddings on each behavior platform, and apply the differential privacy technique to the aggregated user embedding on user server by adding Laplacian noise for perturbation, making it more difficult to infer the raw user behaviors from the local user embeddings and aggregated user embeddings.
Thus, the proposed FedCTR method can well protect user privacy when utilizing user behaviors on different platforms to model user interest for CTR prediction.

\subsubsection{Model Details}\label{section.model}

In this section, we introduce the model details in the FedCTR framework, including the \textit{user model}, \textit{ad model}, \textit{aggregator} and \textit{CTR predictor}.
\begin{figure*}[!t]
  \centering
  \subfigure[User model.]{
    \includegraphics[height=2.0in]{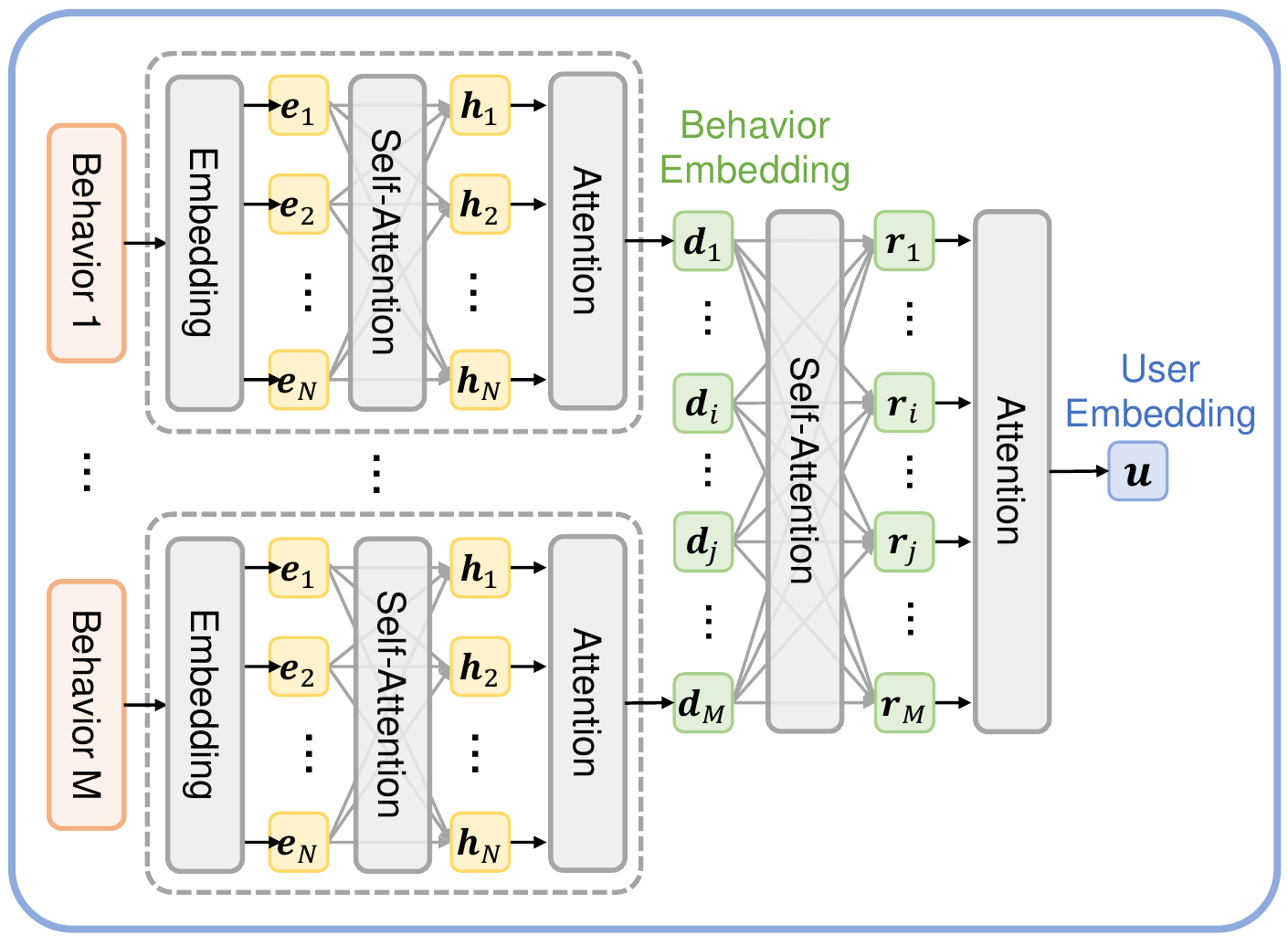}  
  \label{fig.usermodel}
    }\hspace{0.6in}
      \subfigure[Ad model.]{
    \includegraphics[height=2.0in]{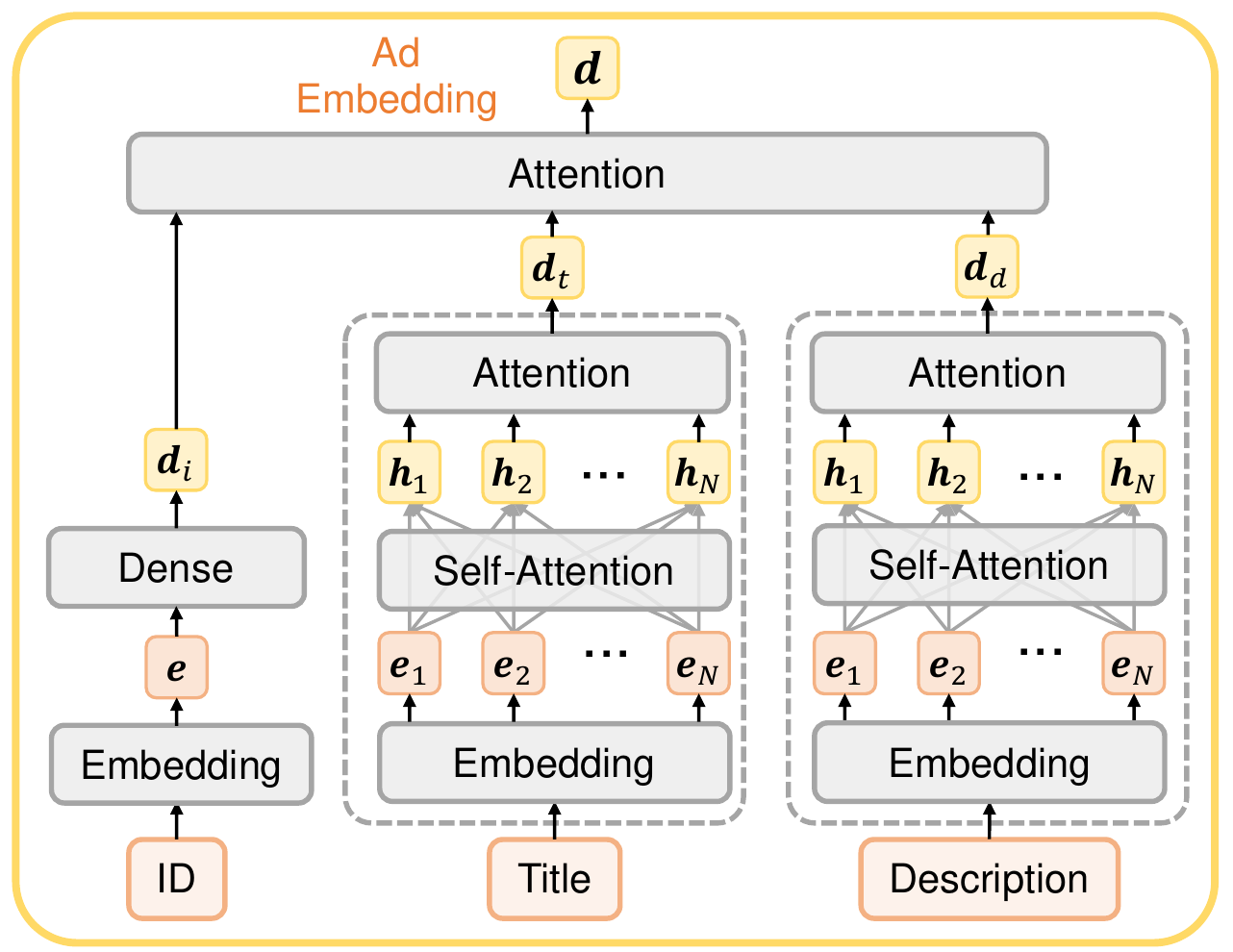}
  \label{fig.admodel}
  }
   \caption{The architecture of the user and ad models.}
\end{figure*}

\textbf{User Model.}\  
User model is used to learn local user embeddings from local user behaviors on the behavior platforms.
The user models on different behavior platforms share the same architecture  but have different model parameters.
The architecture of user model is shown in Fig.~\ref{fig.usermodel}.
It is based on the neural user model proposed in~\cite{wu2019nrms}, which learns user embeddings from user behaviors in a hierarchical way.
It first learns behavior representations from the texts in behaviors, such as the search query in online search behaviors and the webpage title in webpage browsing behaviors.
The behavior representation module first converts the text in behaviors into a sequence of word embeddings.
In addition, following~\cite{devlin2019bert} we add position embedding to each word embedding to model word orders.
Then the behavior representation module uses a multi-head self-attention network~\cite{vaswani2017attention} to learn contextual word representations by capturing the relatedness among the words in the text.
Finally, the behavior representation module applies an attentive pooling network~\cite{yang2016hierarchical} to these contextual word representations which can compute the relative importance of these words and obtain a summarized text representation based on the word representations and their attention weights.

After learning the representations of behaviors, a user representation learning module is used to learn user embedding from these behavior embeddings.
First, we add a position embedding vector to each behavior embedding vector to capture the sequential order of the behaviors.
Then we apply a multi-head self-attention network to learn contextual behavior representations by capturing the relatedness between the behaviors.
Finally, we use an attentive pooling network~\cite{yang2016hierarchical} to obtain a unified user embedding vector by summarizing these contextual behavior representations with their attention weights.
The model parameters of the user model on the $i$-th behavior platform are denoted as $\Theta_{U_i}$, and the learning of local user embedding on this platform can be formulated as $\mathbf{u}_i=f_{user}^i(behaviors; \Theta_{U_i})$.

\textbf{Ad Model.}\ 
The ad model is used to learn embeddings of ads from their IDs, titles, and descriptions.
The architecture of the \textit{ad model} is illustrated in Fig.~\ref{fig.admodel}.
It is based on the ad encoder model proposed in~\cite{an2019neural} with small variants.
Similar with the \textit{user model}, we use a combination of word embedding layer, multi-head self-attention layer and attentive pooling layer to learn the embeddings of titles and descriptions from the texts.
In addition, we use an ID embedding layer and a dense layer to learn ad representation from ad ID.
The final ad representation is learned from the ID embedding, title embedding and description embedding via an attention network~\cite{an2019neural}.
The model parameters of the ad model are denoted as $\Theta_D$, and the learning of ad embedding can be formulated as $\mathbf{d}=f_{ad}(ID, text; \Theta_D)$.

\textbf{Aggregator.}\ 
The aggregator model aims to aggregate the local user embeddings learned from different behavior platforms into a unified user embedding for CTR prediction.
Since user behaviors on different platforms may have different informativeness for modeling user interest, we use an attention network~\cite{yang2016hierarchical} to evaluate the importance of different local user embeddings when synthesizing them together.
It takes the local user embeddings $\mathbf{U}=[\mathbf{u}_1, \mathbf{u}_2, ..., \mathbf{u}_K]$ from the $K$ platforms as the input, and learns the aggregated user embedding $\mathbf{u}$ from them via an attention network, which is formulated as follows:
\begin{equation} 
   \mathbf{u}=f_{agg}(\mathbf{U}; \Theta_A)=\mathbf{U}[\mathrm{softmax}(\mathbf{U}^T\Theta_A)],
\end{equation}
where $\Theta_A$ is the parameters of the aggregator.

\textbf{CTR Predictor.} \ 
The CTR predictor aims to estimate the probability score of a user $u$ clicking a candidate ad $d$ based on their representations $\mathbf{u}$ and $\mathbf{d}$, which is formulated as $\hat{y} = f_{CTR}(\mathbf{u},\mathbf{d}; \Theta_C)$, where $\Theta_C$ is the model parameters of the CTR predictor.
There are many options for the CTR prediction function $f_{CTR}(\cdot)$, such as dot product~\cite{an2019neural}, outer product~\cite{he2018outer} and factorization machine~\cite{guo2017deepfm}.

\subsection{Federated Model Training}\label{section.training}

\begin{figure*}[!t]
  \centering
    \includegraphics[width=0.85\linewidth]{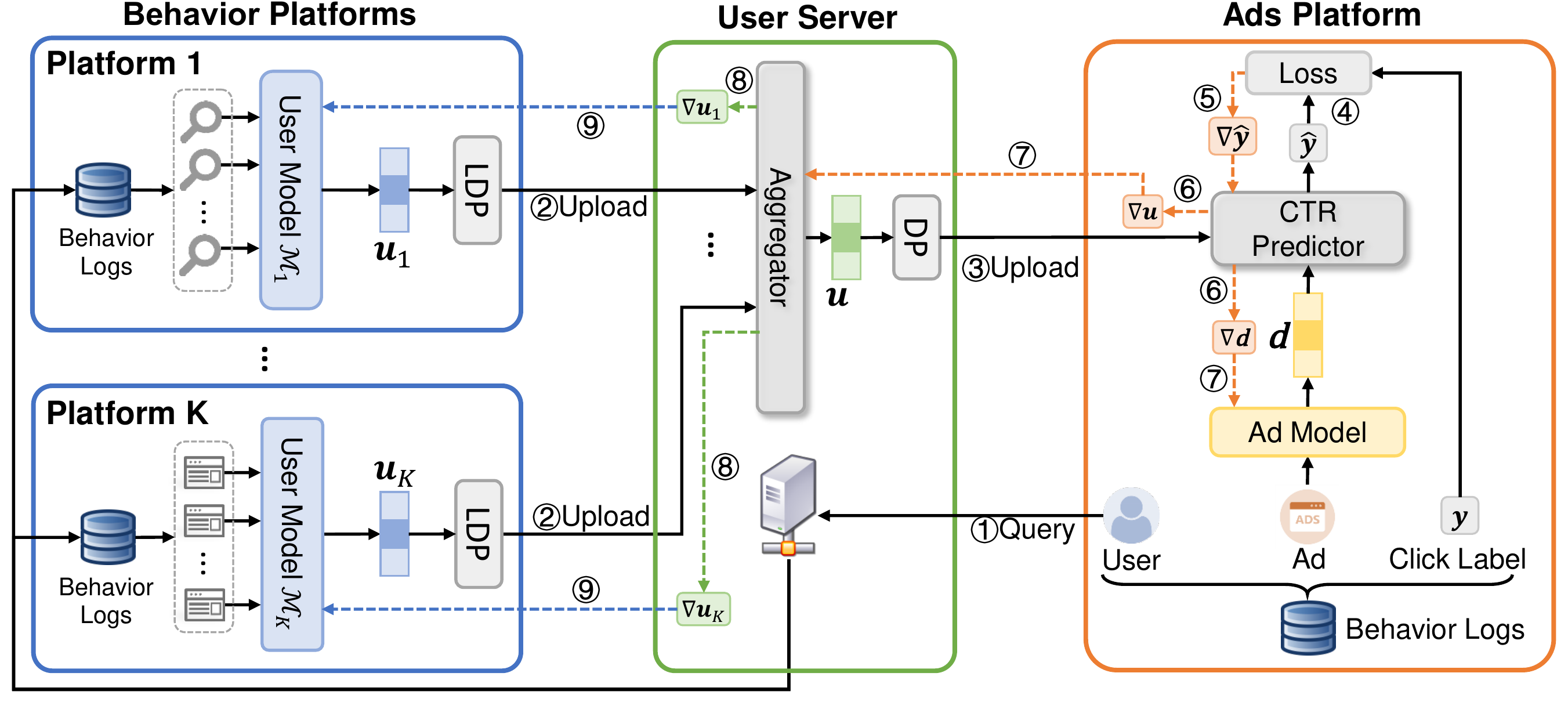}
    \caption{The model training framework of our \textit{FedCTR} approach.}
  \label{fig.model2}
   
\end{figure*}

Existing CTR prediction methods usually train the models in a centralized way, where both the training data and the model parameters are located in the same place.
In the proposed FedCTR method for native ad CTR prediction, the training data is distributed on multiple platforms.
For example, the ad information and the users' click and non-click behaviors on ads which can serve as labels for model training are located in the ad platform, while users' behaviors on many online platforms are located in other behavior platforms.
Due to privacy constraints, the different kinds of user behaviors cannot be centralized.
In addition, FedCTR contains multiple models such as user models, ad model, and aggregator model which are also distributed on different platforms.
Thus, it is a non-trivial task to train the models of FedCTR without violating the privacy protection requirement.
Motivate by~\cite{mcmahan2017communication}, in this section we present a privacy-preserving framework to train the models of FedCTR where each platform learns the model on it in a federated way, and only model gradients (rather than raw user behaviors) are communicated across different platforms.
The framework for model training is shown in Fig.~\ref{fig.model2}.

At the model training stage, for a randomly selected user behavior on the  ad platform denoted as $(u, d, y, t)$ which means at timestamp $t$ an ad $d$ is displayed to user $u$ and her click behavior is $y$ (1 for click and 0 for non-click), we first use the FedCTR framework to learn the local user embeddings $[\mathbf{u}_1, \mathbf{u}_2, ..., \mathbf{u}_K]$ and the aggregated user embedding $\mathbf{u}$ at timestamp $t$ using the current user and aggregator models.
We also use the current ad model to learn an embedding $\mathbf{d}$ of this ad.
Then we use the current CTR predictor model to compute the predicted click probability score $\hat{y}$. 
By comparing $\hat{y}$ with $y$, we can compute the loss of the current FedCTR models on this training sample.
In our model training framework cross-entropy loss is used, and loss of this sample can be formulated as:
\begin{equation} 
    \mathcal{L}(u,d; \Theta_D, \Theta_C, \Theta_A, \Theta_U)=-y\log(\hat{y})-(1-y)\log(1-\hat{y}),
\end{equation}
where $\Theta_U = [\Theta_{U_1},\Theta_{U_2},...,\Theta_{U_K}]$ is the parameter set of user models on different platforms.
% We repeat this process by selecting a batch of training samples, and denote their average loss as $\mathcal{L}$.

Then, we compute the model gradients for model update.
First, according to the loss $\mathcal{L}$, we compute the gradient of $\hat{y}$, denoted as $\nabla\hat{y}$.
Then we input $\nabla\hat{y}$ to the CTR predictor.
Based on $\nabla\hat{y}$ and the CTR prediction function $\hat{y}=f_{CTR}(\mathbf{u},\mathbf{d};\Theta_C)$, we can compute the gradient of parameters $\Theta_C$ in CTR predictor (denoted as $\nabla \Theta_C$), the gradient of user embedding $\mathbf{u}$ (denoted as $\nabla \mathbf{u}$) and the gradient of ad embedding $\mathbf{d}$ (denoted as $\nabla \mathbf{d}$).
The model parameters of the CTR predictor $\Theta_C$ can be updated using $\nabla \Theta_C$ following the SGD~\cite{bottou2010large} algorithm, i.e., $\Theta_C = \Theta_C - \eta \nabla \Theta_C$, where $\eta$ is the learning rate.

The gradient of ad embedding $\nabla \mathbf{d}$ is then sent to the ad model. 
Based on $\nabla \mathbf{d}$ and the function $\mathbf{d}=f_{ad}(ID, text; \Theta_D)$  that summarizes the neural architecture of \textit{ad model}, we can compute the gradient of the ad model, denoted as $\nabla \Theta_D$.
We use $\nabla \Theta_D$ to update the model parameters of ad model $\Theta_D$, i.e., $\Theta_D = \Theta_D - \eta \nabla \Theta_D$.

The user embedding gradient $\nabla\mathbf{u}$ is distributed to the aggregator in the user server.
Based on the aggregator model function $\mathbf{u}=f_{agg}(\mathbf{U}; \Theta_A)$ and $\nabla\mathbf{u}$, we compute the gradients of the aggregator model (denoted as $\nabla \Theta_A$) as well as the gradient of each local user embedding (denoted as $\nabla \mathbf{u}_i, i=1,...,K$).
The model parameters of the aggregator can be updated as $\Theta_A = \Theta_A - \eta \nabla \Theta_A$.
The gradients of the local embeddings, i.e., $\nabla \mathbf{u}_i,i=1,...,K$, are sent to corresponding behavior platform respectively.

On each behavior platform, when it receives the gradient of the local user embedding, the platform will combine the local user embedding gradient, the input user behaviors and the current user model based on the function $\mathbf{u}_i=f_{user}^i(behaviors; \Theta_{U_i})$ to compute the gradient of the user model, which is denoted as $\nabla \Theta_{U_i}$ on the $i$-th platform.
Then the model parameters of the local user models are updated as $\Theta_{U_i} = \Theta_{U_i} - \eta \nabla \Theta_{U_i}$.

Above model training process is conducted on different training samples for multiple rounds until models converge.
In our federated model training framework, the user behaviors on different platforms (e.g., the ad platform and the multiple behavior platforms for user interest modeling) never leave the local platform, and only the model gradients are distributed from the ad platform to user server, and from user server to each user behavior platform.
Since model gradients usually contain much less private information than the raw user behaviors~\cite{mcmahan2017communication}, user privacy can be protected at the training stage of the proposed FedCTR method.

\section{Experiments}\label{sec:Experiments}

\subsection{Datasets and Experimental Settings}

Since there is no publicly available dataset for native ad CTR prediction, we constructed one by collecting the logs of 100,000 users on the native ads displayed on a commercial platform  from 11/06/2019 to 02/06/2020.%\footnote{The platform name is anonymized for double-blind review.}
The logs in the last week were reserved for testing, and the remaining logs were used for model training.
We randomly selected 10\% of the samples in training set for validation.
We also collected the search logs and webpage browsing behavior logs of  users recorded by a commercial search engine during the same period for user interest modeling.
We assume that different kinds of user behavior data come from independent platforms and they cannot be directly aggregated.
The detailed statistics of the dataset are shown in Table~\ref{dataset}.

\begin{table}[h]
\centering
\caption{Detailed statistics of the dataset for native ad CTR prediction.}\label{dataset}
\resizebox{0.48\textwidth}{!}{
\begin{tabular}{lrlr}
\Xhline{1.5pt}
\#users                    & 100,000   & avg. \#words per ad title       & 3.73   \\
\#ads                      & 8,105     & avg. \#words per ad description & 15.31  \\
\#ad click behaviors       & 345,264   & avg. \#words per search query   & 4.64   \\
\#ad non-click behaviors   & 345,264   & avg. \#words per webpage title  & 10.84  \\
avg. \#queries per user    & 50.69     & avg. \#webpages per user        & 210.09 \\ \Xhline{1.5pt}
\end{tabular}
}

\end{table}

In our experiments, the word embeddings in the user and ad models were initialized by the pre-trained Glove~\cite{pennington2014glove} embeddings.
In the CTR predictor, following~\cite{an2019neural} we used dot-product as the CTR prediction function.
%Each self-attention networks had 16 heads, and the output from each head was 16-dimensional.
%The query vectors in attention networks were 256-dimensional.
% Adam~\cite{kingma2014adam} was used as the model optimizer ($\eta$=1e-3).
% The batch size was 32.
The dropout~\cite{srivastava2014dropout} ratio after each layer was 20\%.
The strength of Laplace noise $\lambda_{LDP}$ in the LDP modules was 0.01, and $\lambda_{DP}$ in the DP module was 0.005.
Hyperparamters were tuned on the validation set.
To evaluate the model performance, on the \textit{Ads} dataset we used AUC and AP as the metrics.
We reported the average results of 10 independent experiments.

\subsection{CTR Prediction Performance}

We compare the performance of \textit{FedCTR} with several baseline methods, include:
\begin{itemize}
    \item LR~\cite{chakrabarti2008contextual,richardson2007predicting}, logistic regression, a widely used method for ads CTR prediction. We used ad IDs and  the TF-IDF features extracted from the texts of behaviors and ads as the input.
   \item LibFM~\cite{rendle2012factorization}, a popular feature-based factorization machine for CTR prediction. We used the same features as LR.
   \item Wide\&Deep~\cite{cheng2016wide}, a popular CTR prediction method with a wide linear part and a deep neural part. Same features with LR were used.
   \item PNN~\cite{qu2016product}, product-based neural network for CTR prediction which can model the interactions between features.
   \item DSSM~\cite{huang2013learning}, deep structured semantic model, a famous model for CTR prediction and recommendation.
   \item DeepFM~\cite{guo2017deepfm}, a combination of factorization machines and deep neural networks for CTR prediction.
   \item NativeCTR~\cite{an2019neural}, a neural native ads CTR prediction method based on attentive multi-view learning.
\end{itemize}

\begin{table}[!t]
 \caption{Comparisons of different methods. * means the ideal performance under centralized user behavior data from different platforms.} \label{table.performance} 
\resizebox{0.48\textwidth}{!}{
\begin{tabular}{lcccccc}
\Xhline{1.5pt}
\multirow{2}{*}{\textbf{Methods}}  & \multicolumn{2}{c}{25\%} & \multicolumn{2}{c}{50\%} & \multicolumn{2}{c}{100\%} \\ \cline{2-7} 
                                                & AUC         & AP         & AUC         & AP         & AUC         & AP          \\ \hline
LR\cite{richardson2007predicting}                                       & 58.42       & 56.47      & 58.96       & 56.87      & 59.36       & 57.24       \\
LR*                                           & 60.82       & 57.38      & 61.44       & 58.60      & 62.04       & 59.20       \\ \hline
LibFM\cite{rendle2012factorization}                                      & 57.79       & 55.91      & 58.10       & 56.33      & 58.45       & 56.71       \\
LibFM*                                     & 61.59       & 58.13      & 61.91       & 59.17      & 62.47       & 59.99       \\ \hline
Wide\&Deep\cite{cheng2016wide}                                 & 59.45       & 57.80      & 59.66       & 57.97      & 60.04       & 58.61       \\
Wide\&Deep*                                         & 62.10       & 59.75      & 62.35       & 59.87      & 62.79       & 60.28       \\ \hline
PNN\cite{qu2016product}                                      & 59.53       & 57.86      & 59.73       & 58.03      & 60.02       & 58.50       \\
PNN*                                                  & 62.54       & 60.12      & 62.73       & 60.29      & 62.87       & 60.41       \\ \hline
DSSM\cite{huang2013learning}                                     & 59.24       & 57.59      & 59.46       & 57.89      & 59.92       & 58.43       \\
DSSM*                                               & 62.23       & 59.66      & 62.50       & 59.85      & 62.85       & 60.37       \\ \hline
DeepFM\cite{guo2017deepfm}                                   & 59.36       & 57.70      & 59.55       & 57.92      & 59.83       & 58.28       \\
DeepFM*                                         & 61.88       & 59.47      & 62.05       & 59.77      & 62.72       & 60.24       \\ \hline
NativeCTR\cite{an2019neural}                                & 60.84       & 59.12      & 61.12       & 59.40      & 61.44       & 59.75       \\
NativeCTR*                           & 62.88       & 60.69      & 63.01       & 60.88      & 63.39       & 61.17       \\ \hline
FedCTR                                  & \textbf{63.95}       & \textbf{61.82}      & \textbf{64.20}       & \textbf{62.13}      & \textbf{64.54}       & \textbf{62.50}       \\
%FedCTR**                                & 64.31       & 62.33      & 64.61       & 62.69      & 64.89       & 63.01       \\ \hline
 \Xhline{1.5pt}
\end{tabular}
}

\end{table}

We report the performance of baseline methods based on user behaviors on the ads platform only and their ideal performance using the centralized storage of behavior data from different platforms.
The results under different ratios of training data of different methods are shown in Table~\ref{table.performance}.
According to the results, we find the methods using neural networks to learn user and ad representations (e.g., \textit{FedCTR}) perform better than those using handcrafted features to represent users and ads (e.g., \textit{LR} and \textit{LibFM}).
It shows that the representations learned by neural networks are more suitable than handcrafted features in modeling users and ads.
In addition, compared with the methods solely based on ad click behaviors on the ad platform for user interest modeling, the methods that consider multi-platform behaviors in user modeling can achieve better performance.
This is because the user behavior data on the ad platform may be sparse, which is insufficient to infer user interest accurately.
Since user behaviors on different platforms can provide rich clues for inferring user interest in different aspects, incorporating multi-platform user behaviors is beneficial for user interest modeling.
Unfortunately, user behavior data is highly privacy-sensitive and usually cannot be centrally stored or exchanged among platforms due to the constraints of data protection regulations like GDPR and the privacy concerns from users.
Thus, in many situations the methods that need centralized storage of multi-platform user behavior data may not achieve the ideal performance in Table~\ref{table.performance}.
Besides, our \textit{FedCTR} method consistently outperforms other baseline methods.
This is because our framework is more effective in leveraging multi-platform user behaviors for user interest modeling than other baseline methods, and the ad and user models also have greater ability in learning ad and user representations. 
Moreover, in our approach the raw behavior data never leaves the local platforms, and only user embeddings and model gradients are communicated among different platforms.
Thus, our approach can leverage multi-platform user behavior data for user interest modeling in a privacy-preserving manner.
%Besides, since the local and aggregated user embedding may contain some private information on user behaviors,  we add LDP and DP techniques respectively to the local and aggregated user embeddings for better  privacy protection.
%Compared with the performance of  \textit{FedCTR} and its variant  without LDP and DP, we find the performance decline brought by LDP and DP is small.
%Thus, our approach can protect user privacy with acceptable performance loss.

\begin{figure}[!t]
  \centering
  \subfigure[Influence of the number of behavior platform for user modeling.]{
    \includegraphics[height=1.8in]{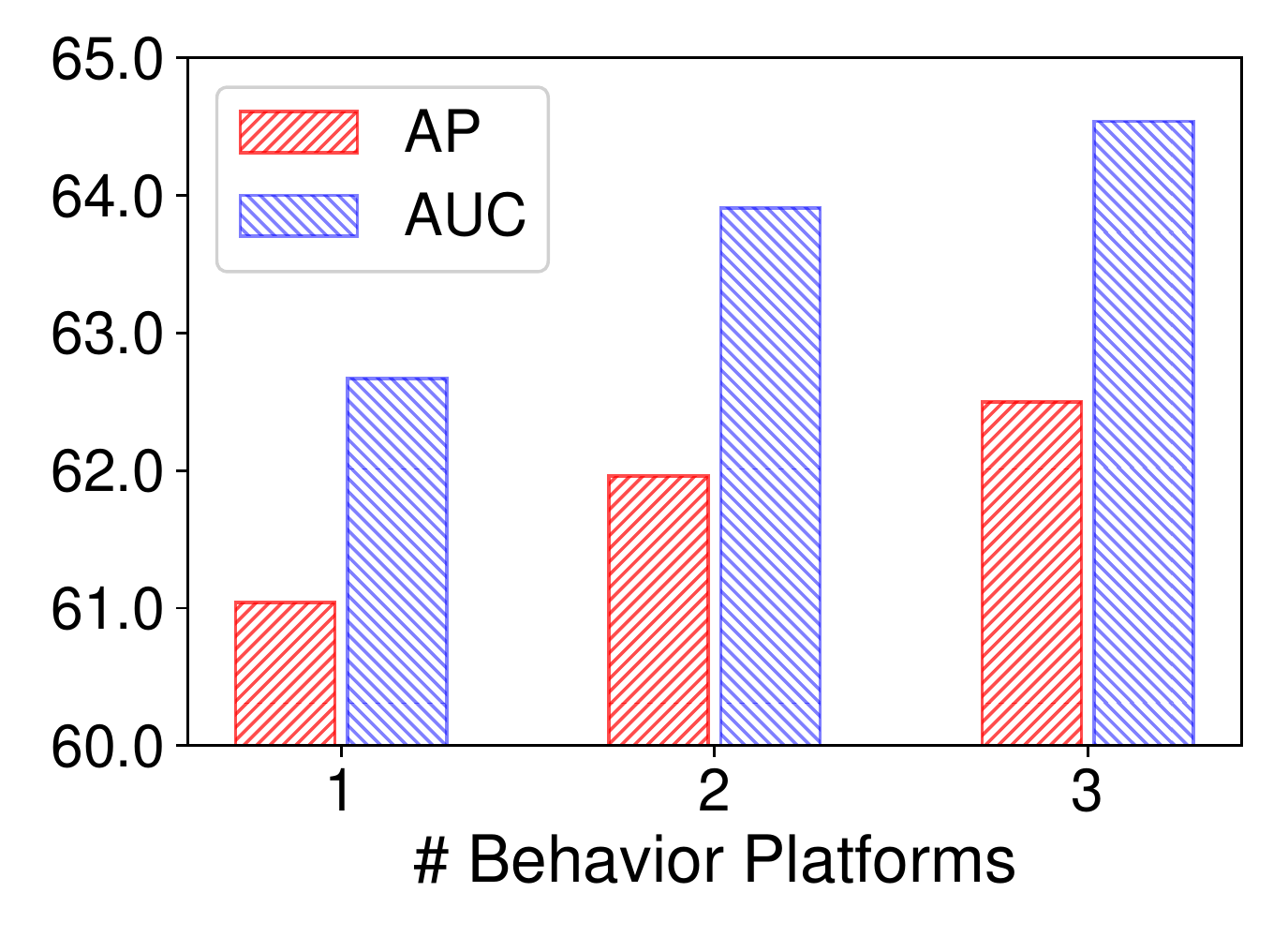}
  \label{fig.behavior}
  }
   \subfigure[Influence of the number of user behaviors.]{
      \includegraphics[height=1.8in]{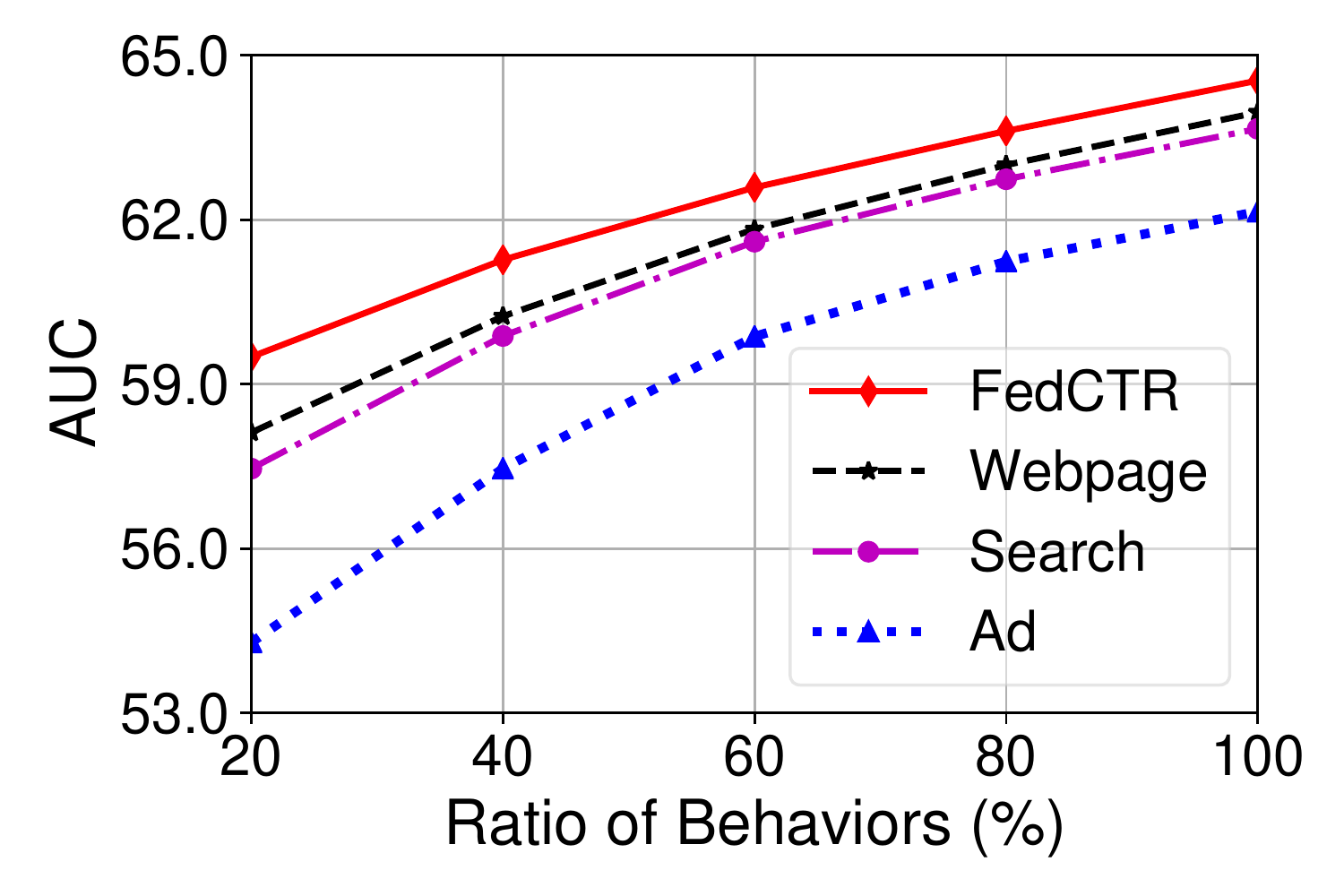}
  \label{fig.behaviornum}
  }
  \caption{Effect of multi-platform user behaviors.}
\end{figure}

\begin{figure}[!t]
  \centering

\end{figure}

\subsection{Effect of Multi-Platform Behaviors}

Next, we explore the effectiveness of incorporating user behavior data from multiple platforms for CTR prediction.
We first study the influence of the number of behavior platforms for user modeling.
The average results of \textit{FedCTR} with different numbers of platforms are shown in Fig.~\ref{fig.behavior}.
From the results, we find that the performance improves as number of platforms increases.
This is probably because user behaviors on different platforms can provide complementary information to help cover user interest more comprehensively.
It shows that incorporating multi-platform user behaviors can effectively enhance user interest modeling for CTR prediction.

We also study the influence of the number of user behaviors on each platform on the CTR prediction performance.
We vary the ratios of user behaviors for user interest modeling from 20\% to 100\%, and the results are shown in Fig.~\ref{fig.behaviornum}.
From the results, we find that the performance of \textit{FedCTR} declines when the number of behaviors decreases.
It indicates that it is more difficult to infer user interest when user behaviors are scarce.
In addition, we find that the \textit{FedCTR} method consistently outperforms its variants with single-platform user behaviors, and the advantage becomes larger when user behaviors are scarcer.
It shows that utilizing the user behavior data decentralized in different platforms can help model the interest of users more accurately, especially when user behaviors on a single platform are sparse.

\begin{figure}[!t]
  \centering
  \subfigure[Local embedding.]{
    \includegraphics[width=0.38\textwidth]{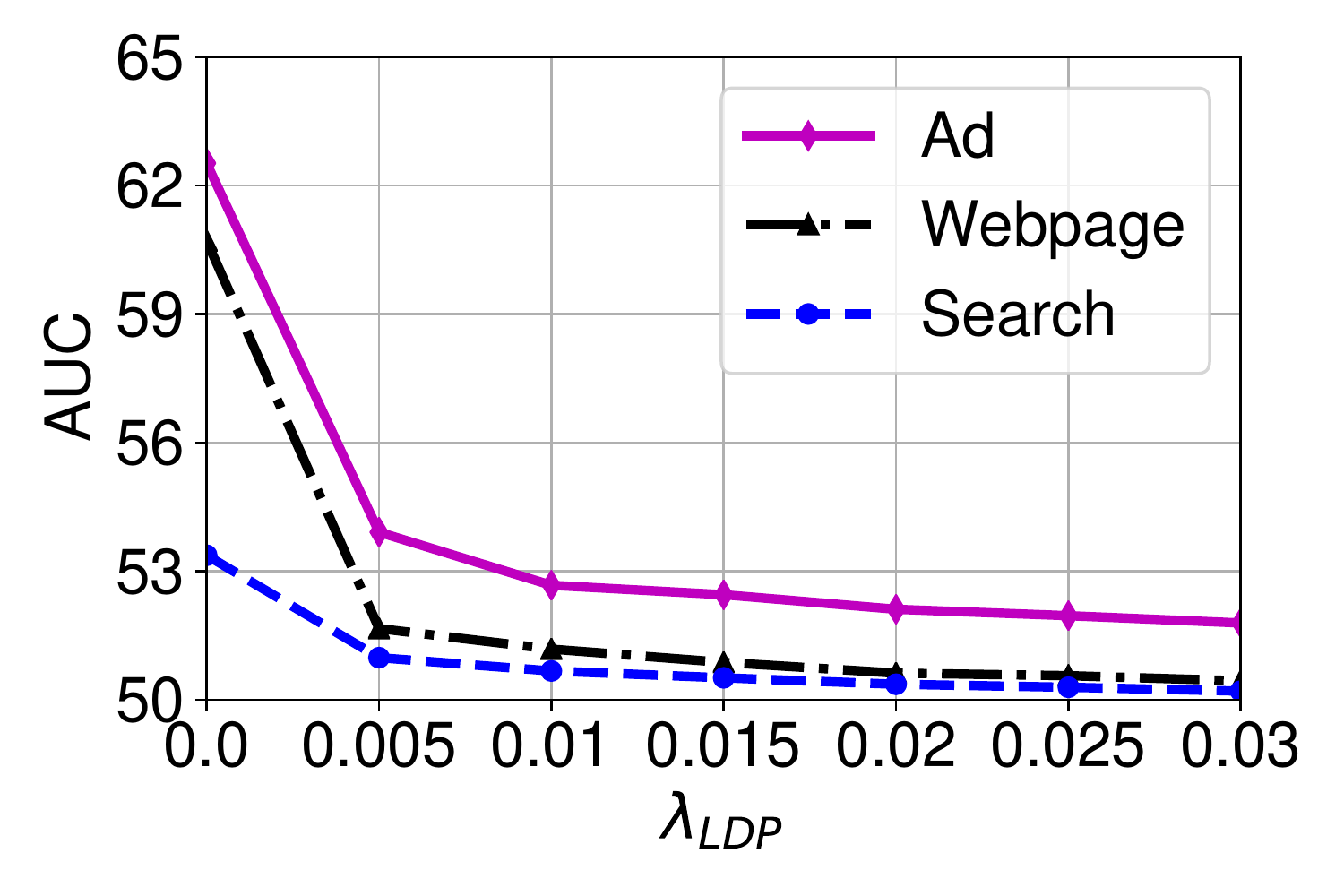}  \label{fig.attack1}
    }
    \subfigure[Aggregated embedding.]{
  \includegraphics[width=0.38\textwidth]{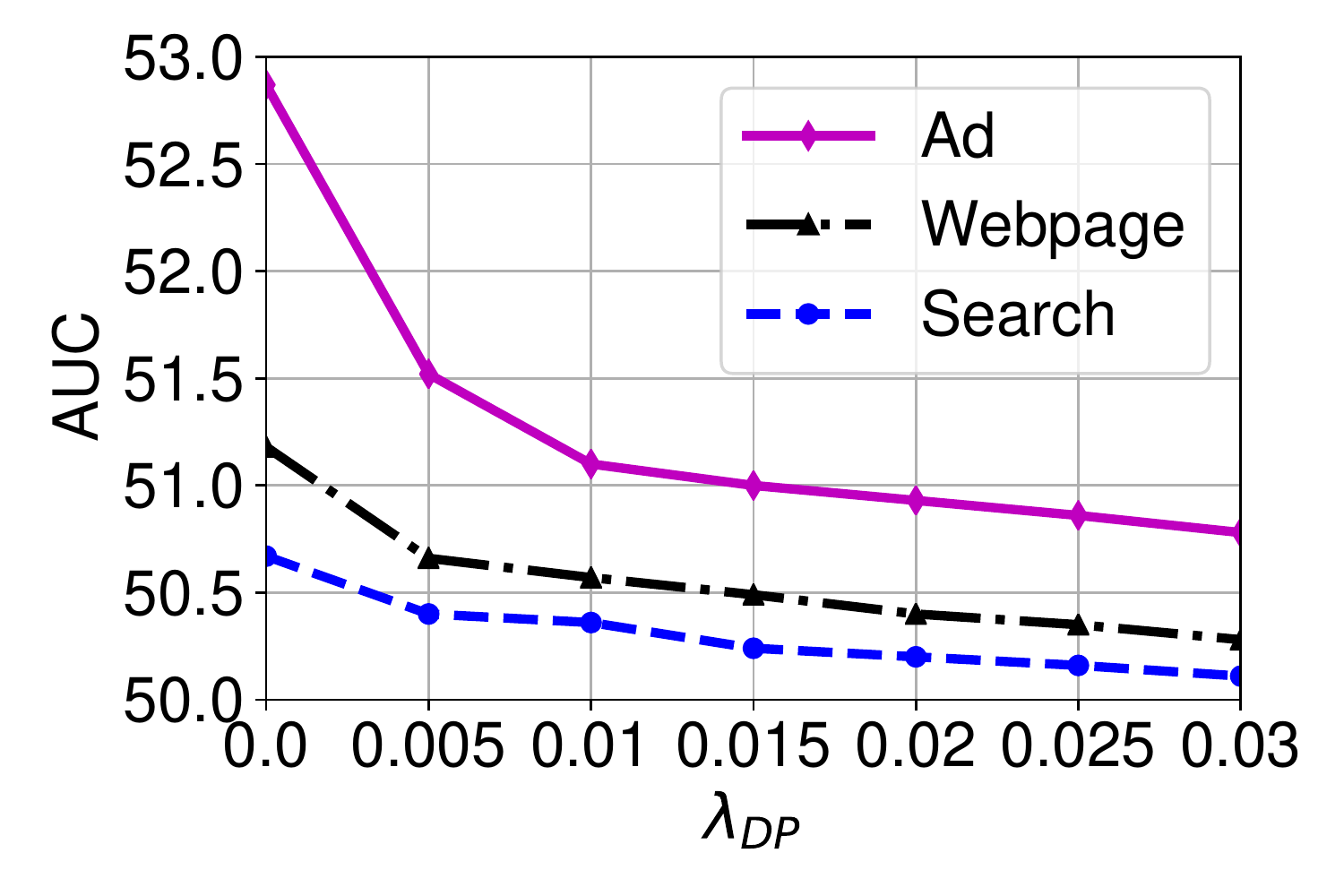}  \label{fig.attack2}
    }   
  \caption{Privacy protection of local and aggregated user embedding under different $\lambda_{LDP}$ and $\lambda_{DP}$ values. Lower AUC indicates better privacy protection.} 
\end{figure}

\begin{figure}[!t]
  \centering
  \subfigure[$\lambda_{LDP}$.]{
    \includegraphics[width=0.35\textwidth]{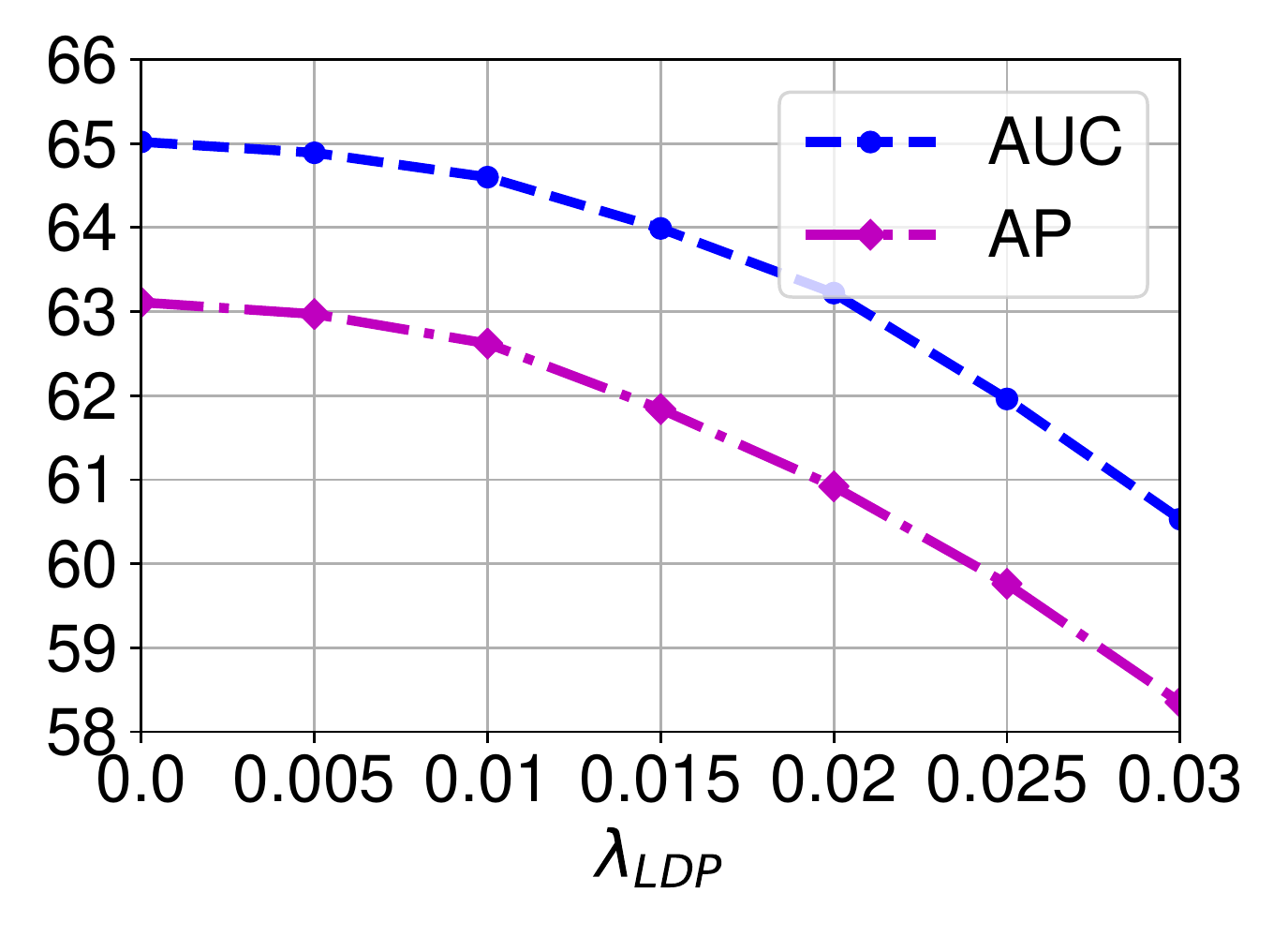}  \label{fig.auc1}
    } 
    \subfigure[$\lambda_{DP}$ .]{
  \includegraphics[width=0.35\textwidth]{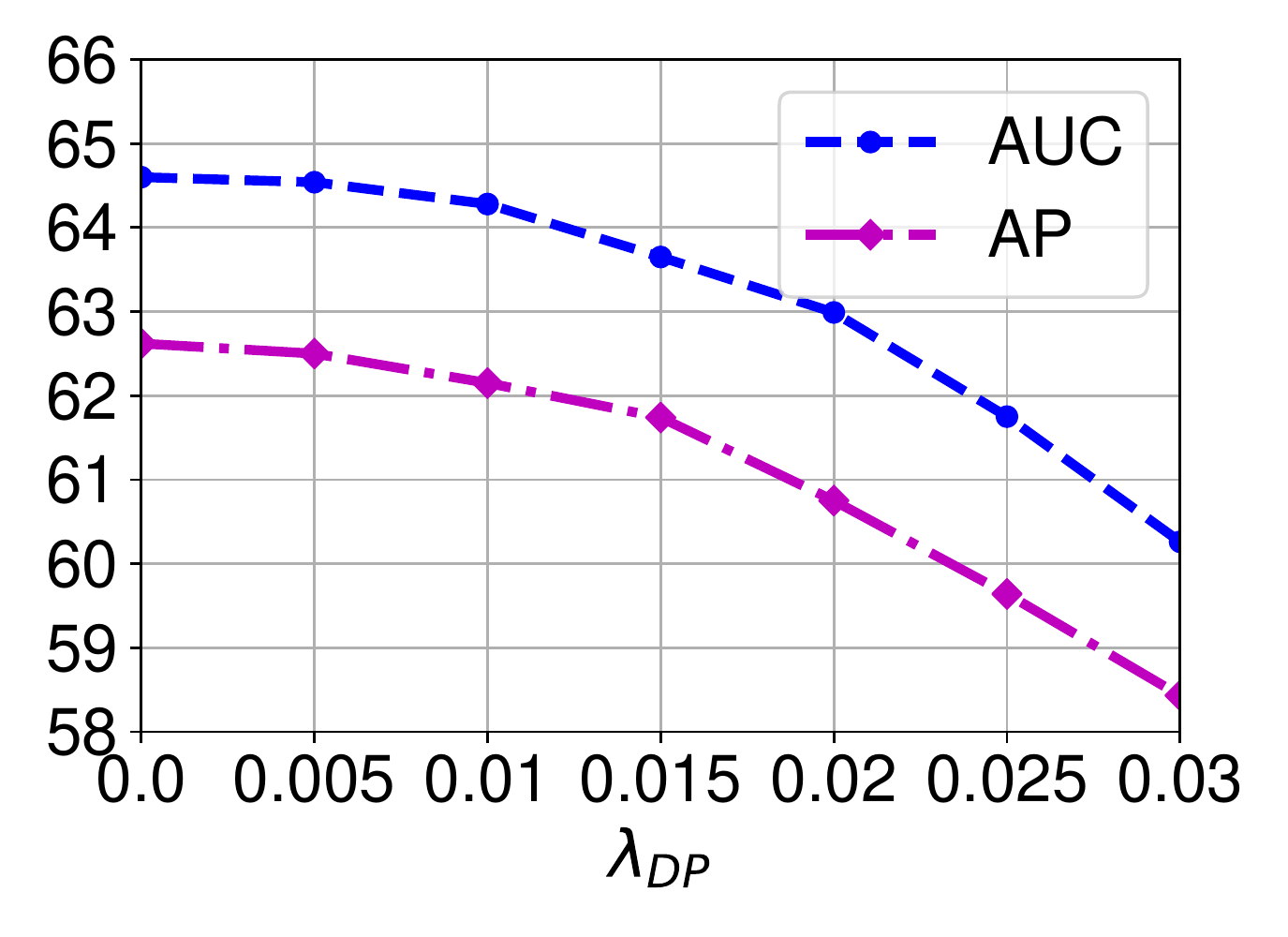}  \label{fig.auc2}
    }
  \caption{Influence of $\lambda_{LDP}$ and $\lambda_{DP}$ on CTR prediction.}
\end{figure}

\subsection{Study on Privacy Protection}
 
In this section, we verify the effectiveness of our \textit{FedCTR} method in privacy protection when exploiting multi-platform user behavior data for user modeling.
Since user embeddings learned from different platforms may contain private information that can be used to infer the raw behavior data, we use LDP and DP to protect the local and aggregated user embeddings, respectively.
To quantitatively evaluate the privacy protection performance of these embeddings, we use a behavior prediction task by predicting the raw user behaviors from the local and aggregated user embeddings to indicate their privacy protection ability.
More specifically, for each user we randomly sample a real behavior of this user (regarded as a positive sample) and 9 behaviors which do not belong to this user (regarded as negative samples).
The goal is to infer the real behavior from these 10 candidate behaviors by measuring their similarities to the user embedding.
We perform dot product between the embeddings of each user-behavior pair, then all candidate behaviors are ranked according to the computed scores.
We use AUC as the metric to evaluate the privacy protection performance, and lower AUC scores indicate better privacy protection.

There are two key hyperparameters in the LDP and DP modules, i.e., $\lambda_{LDP}$ and $\lambda_{DP}$, which control the strength of the Laplacian noise added to user embeddings.
We first vary the value of $\lambda_{LDP}$ to explore its influence on  privacy protection and CTR prediction, and the results are shown in Figs.~\ref{fig.attack1} and  \ref{fig.auc1}, respectively.\footnote{The DP module is deactivated in these experiments.}
We find that although the CTR prediction performance is slightly better if the local user embeddings are not protected by LDP, the raw user behaviors can be inferred to a certain extent, which indicates that the private information of local user embeddings is not fully protected.
Thus, it is important to use LDP to protect the local embeddings learned from different behavior platforms.
In addition, if $\lambda_{LDP}$ is too large, the CTR performance declines significantly, and the improvement on privacy protection is marginal.
Thus, we choose a moderate $\lambda_{LDP}$ (i.e., 0.01) to achieve a trade-off between CTR prediction and privacy protection.
Then, we explore the influence of $\lambda_{DP}$ on the performance of \textit{FedCTR} in privacy protection and CTR prediction (under $\lambda_{LDP}=0.01$), and the results  are respectively shown in Figs.~\ref{fig.attack2} and \ref{fig.auc2}. 
We find it is also important to set a moderate value for $\lambda_{DP}$ (e.g., 0.005) to balance the performance of CTR prediction and privacy protection.
Besides, comparing the privacy protecting performance on local and aggregated embeddings, we find that it is more difficult to infer the raw user behaviors on a specific platform from the aggregated user embedding than from local user embeddings.
This may be because the aggregated user embedding is a summarization of local embeddings, and the private information is more difficult to be recovered.

\subsection{Effect of Aggregator and CTR Predictor}\label{exp.model}
 
\begin{figure}[!t]
  \centering
  \subfigure[CTR predictor.]{\label{fig.predictor}
    \includegraphics[width=0.4\textwidth]{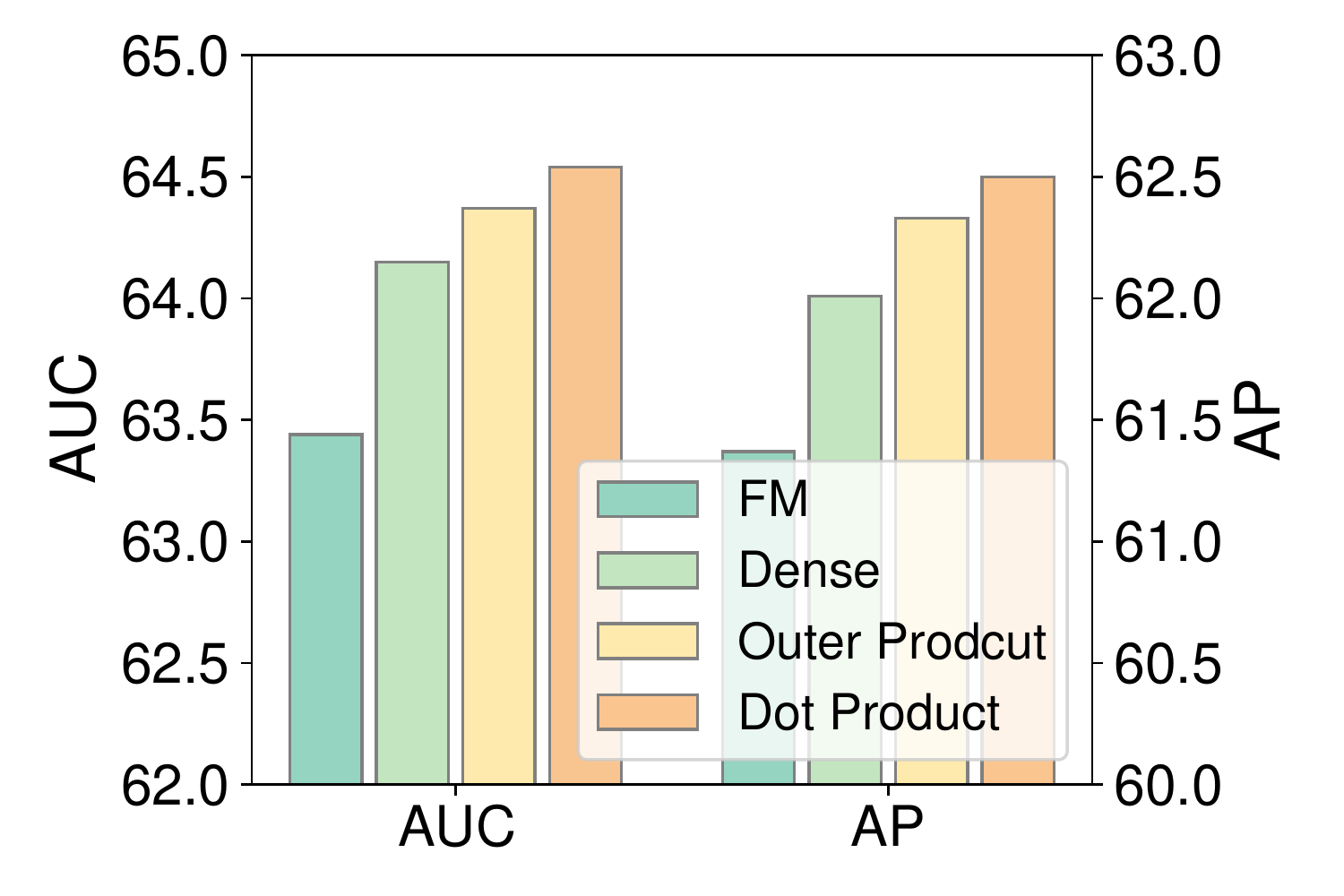}
  }
    \subfigure[Aggregator.]{ \label{fig.aggregator}
      \includegraphics[width=0.4\textwidth]{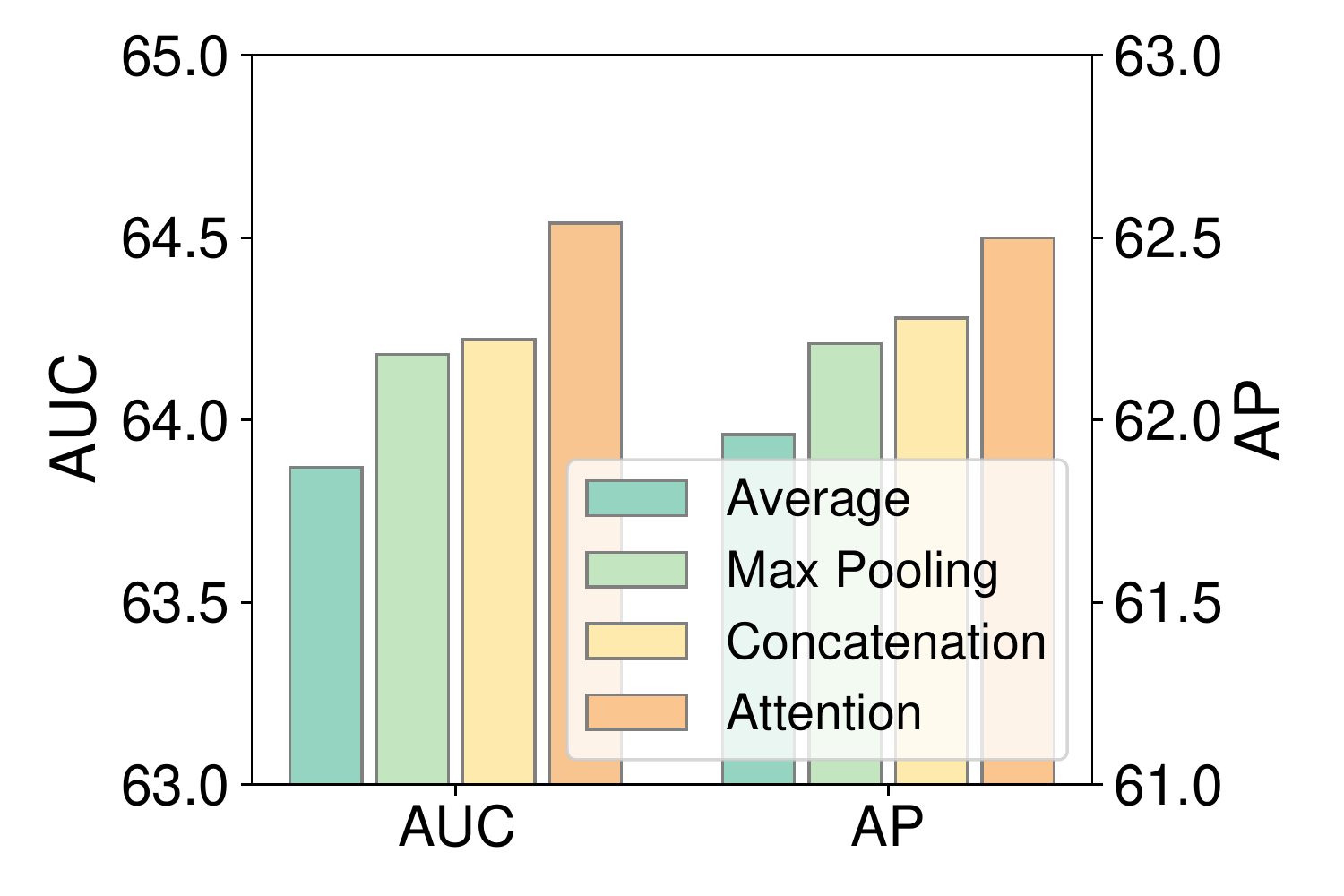}
  }
  \caption{Different  CTR predictor and aggregator models.}  
\end{figure}

We also verify the effectiveness of the aggregator and CTR predictor models.
First, we compare different CTR prediction models like factorization machine (FM)~\cite{guo2017deepfm}, dense layers, outer product~\cite{he2018outer} and dot product~\cite{an2019neural}, and the results are shown in Fig.~\ref{fig.predictor}.
From Fig.~\ref{fig.predictor}, we find the performance of FM is not optimal.
It shows that FM may not be suitable for modeling the similarity between the user and ad representations learned by neural networks.
In addition, we find that using a dense layer is also not optimal.
This may be because dense layers compute the click scores based on the concatenation of ad and user representations while difficult to model their interactions, which is also validated by~\cite{rendle2020neural}.
Besides, it is interesting that dot product achieves the best performance.
This may be because dot product simultaneously models the distance between two vectors as well as their lengths, which can effectively measure the relevance of user and ad representations for CTR prediction.
Thus, we prefer dot product for its effectiveness and simplicity.

Then, we compare different models for user embedding aggregation, including attention network, average pooling, max pooling and concatenation.
The results are shown in Fig.~\ref{fig.aggregator}. 
We find that average pooling is sub-optimal for aggregation, since it cannot distinguish the informativeness of different local user embeddings.
In addition, max pooling is also not optimal, since it only keeps the most salient features.
Moreover, although concatenating user embeddings can keep more information, it is inferior to using attention mechanism due to its lack of informativeness modeling.
Thus, we use attention networks to implement the aggregator in the user server.
\section{Conclusion and Future Works}\label{sec:Conclusion}

In this paper, we propose a federated CTR prediction method which can model user interest from user beahaviors on different platforms in a privacy-preserving way.
In our method we learn user embeddings from the multi-platform user behavior data in a federated way.
Each platform learns local user embeddings from the local user behavior data, and upload them to a user server for aggregation.
The aggregated user embedding is sent to the ad platform for CTR prediction.
In addition, we apply LDP and DP techniques to the local and aggregated user embeddings respectively to better protect user privacy.
Besides, we propose a federated model training framework to coordinate different platforms to collaboratively train the models of FedCTR by sharing model gradients rather than raw behaviors to protect user privacy at the model training stage.
Experiments on real-world datasets show that our method is effective in native ad CTR prediction by incorporating multi-platform user behavior data especially when user behaviors on the ad platform are scarce, and meanwhile can effectively protect user privacy.

In future, we plan to deploy FedCTR to online ad system and test its online performance.
We are also interested in applying FedCTR to enhance other CTR prediction tasks (e.g., search ads) by improving their user modeling part via incorporating multi-platform user behavior data in a privacy-preserving way.

\bibliographystyle{ACM-Reference-Format}
\bibliography{main}


\begin{thebibliography}{39}

%%% the Web address.

\ifx \showCODEN    \undefined \def \showCODEN     #1{\unskip}     \fi
\ifx \showDOI      \undefined \def \showDOI       #1{#1}\fi
\ifx \showISBNx    \undefined \def \showISBNx     #1{\unskip}     \fi
\ifx \showISBNxiii \undefined \def \showISBNxiii  #1{\unskip}     \fi
\ifx \showISSN     \undefined \def \showISSN      #1{\unskip}     \fi
\ifx \showLCCN     \undefined \def \showLCCN      #1{\unskip}     \fi
\ifx \shownote     \undefined \def \shownote      #1{#1}          \fi
\ifx \showarticletitle \undefined \def \showarticletitle #1{#1}   \fi
\ifx \showURL      \undefined \def \showURL       {\relax}        \fi
% invisible to TeX
\providecommand\bibfield[2]{#2}
\providecommand\bibinfo[2]{#2}
\providecommand\natexlab[1]{#1}
\providecommand\showeprint[2][]{arXiv:#2}

\bibitem[\protect\citeauthoryear{An, Wu, Wang, Di, Huang, and Xie}{An
  et~al\mbox{.}}{2019}]%
        {an2019neural}
\bibfield{author}{\bibinfo{person}{Mingxiao An}, \bibinfo{person}{Fangzhao Wu},
  \bibinfo{person}{Heyuan Wang}, \bibinfo{person}{Tao Di},
  \bibinfo{person}{Jianqiang Huang}, {and} \bibinfo{person}{Xing Xie}.}
  \bibinfo{year}{2019}\natexlab{}.
\newblock \showarticletitle{Neural CTR Prediction for Native Ad}. In
  \bibinfo{booktitle}{\emph{CCL}}. Springer, \bibinfo{pages}{600--612}.
\newblock


\bibitem[\protect\citeauthoryear{Bottou}{Bottou}{2010}]%
        {bottou2010large}
\bibfield{author}{\bibinfo{person}{L{\'e}on Bottou}.}
  \bibinfo{year}{2010}\natexlab{}.
\newblock \showarticletitle{Large-scale machine learning with stochastic
  gradient descent}.
\newblock In \bibinfo{booktitle}{\emph{COMPSTAT}}.
  \bibinfo{publisher}{Springer}, \bibinfo{pages}{177--186}.
\newblock


\bibitem[\protect\citeauthoryear{Chakrabarti, Agarwal, and
  Josifovski}{Chakrabarti et~al\mbox{.}}{2008}]%
        {chakrabarti2008contextual}
\bibfield{author}{\bibinfo{person}{Deepayan Chakrabarti},
  \bibinfo{person}{Deepak Agarwal}, {and} \bibinfo{person}{Vanja Josifovski}.}
  \bibinfo{year}{2008}\natexlab{}.
\newblock \showarticletitle{Contextual advertising by combining relevance with
  click feedback}. In \bibinfo{booktitle}{\emph{WWW}}.
  \bibinfo{pages}{417--426}.
\newblock


\bibitem[\protect\citeauthoryear{Chen, Sun, Li, Lu, and Hua}{Chen
  et~al\mbox{.}}{2016}]%
        {chen2016deep}
\bibfield{author}{\bibinfo{person}{Junxuan Chen}, \bibinfo{person}{Baigui Sun},
  \bibinfo{person}{Hao Li}, \bibinfo{person}{Hongtao Lu}, {and}
  \bibinfo{person}{Xian-Sheng Hua}.} \bibinfo{year}{2016}\natexlab{}.
\newblock \showarticletitle{Deep ctr prediction in display advertising}. In
  \bibinfo{booktitle}{\emph{MM}}. \bibinfo{pages}{811--820}.
\newblock


\bibitem[\protect\citeauthoryear{Cheng, Koc, Harmsen, Shaked, Chandra, Aradhye,
  Anderson, Corrado, Chai, Ispir, et~al\mbox{.}}{Cheng et~al\mbox{.}}{2016}]%
        {cheng2016wide}
\bibfield{author}{\bibinfo{person}{Heng-Tze Cheng}, \bibinfo{person}{Levent
  Koc}, \bibinfo{person}{Jeremiah Harmsen}, \bibinfo{person}{Tal Shaked},
  \bibinfo{person}{Tushar Chandra}, \bibinfo{person}{Hrishi Aradhye},
  \bibinfo{person}{Glen Anderson}, \bibinfo{person}{Greg Corrado},
  \bibinfo{person}{Wei Chai}, \bibinfo{person}{Mustafa Ispir}, {et~al\mbox{.}}}
  \bibinfo{year}{2016}\natexlab{}.
\newblock \showarticletitle{Wide \& deep learning for recommender systems}. In
  \bibinfo{booktitle}{\emph{DLRS}}. ACM, \bibinfo{pages}{7--10}.
\newblock


\bibitem[\protect\citeauthoryear{Devlin, Chang, Lee, and Toutanova}{Devlin
  et~al\mbox{.}}{2019}]%
        {devlin2019bert}
\bibfield{author}{\bibinfo{person}{Jacob Devlin}, \bibinfo{person}{Ming-Wei
  Chang}, \bibinfo{person}{Kenton Lee}, {and} \bibinfo{person}{Kristina
  Toutanova}.} \bibinfo{year}{2019}\natexlab{}.
\newblock \showarticletitle{BERT: Pre-training of Deep Bidirectional
  Transformers for Language Understanding}. In
  \bibinfo{booktitle}{\emph{NAACL-HLT}}. \bibinfo{pages}{4171--4186}.
\newblock


\bibitem[\protect\citeauthoryear{Dwork}{Dwork}{2008}]%
        {dwork2008differential}
\bibfield{author}{\bibinfo{person}{Cynthia Dwork}.}
  \bibinfo{year}{2008}\natexlab{}.
\newblock \showarticletitle{Differential privacy: A survey of results}. In
  \bibinfo{booktitle}{\emph{TAMC}}. Springer, \bibinfo{pages}{1--19}.
\newblock


\bibitem[\protect\citeauthoryear{Feng and Yu}{Feng and Yu}{2020}]%
        {feng2020multi}
\bibfield{author}{\bibinfo{person}{Siwei Feng} {and} \bibinfo{person}{Han Yu}.}
  \bibinfo{year}{2020}\natexlab{}.
\newblock \showarticletitle{Multi-Participant Multi-Class Vertical Federated
  Learning}.
\newblock \bibinfo{journal}{\emph{arXiv preprint arXiv:2001.11154}}
  (\bibinfo{year}{2020}).
\newblock


\bibitem[\protect\citeauthoryear{Feng, Lv, Shen, Wang, Sun, Zhu, and Yang}{Feng
  et~al\mbox{.}}{2019}]%
        {feng2019deep}
\bibfield{author}{\bibinfo{person}{Yufei Feng}, \bibinfo{person}{Fuyu Lv},
  \bibinfo{person}{Weichen Shen}, \bibinfo{person}{Menghan Wang},
  \bibinfo{person}{Fei Sun}, \bibinfo{person}{Yu Zhu}, {and}
  \bibinfo{person}{Keping Yang}.} \bibinfo{year}{2019}\natexlab{}.
\newblock \showarticletitle{Deep session interest network for click-through
  rate prediction}. In \bibinfo{booktitle}{\emph{AAAI}}.
  \bibinfo{pages}{2301--2307}.
\newblock


\bibitem[\protect\citeauthoryear{Guo, Tang, Ye, Li, and He}{Guo
  et~al\mbox{.}}{2017}]%
        {guo2017deepfm}
\bibfield{author}{\bibinfo{person}{Huifeng Guo}, \bibinfo{person}{Ruiming
  Tang}, \bibinfo{person}{Yunming Ye}, \bibinfo{person}{Zhenguo Li}, {and}
  \bibinfo{person}{Xiuqiang He}.} \bibinfo{year}{2017}\natexlab{}.
\newblock \showarticletitle{DeepFM: a factorization-machine based neural
  network for CTR prediction}. In \bibinfo{booktitle}{\emph{AAAI}}.
  \bibinfo{pages}{1725--1731}.
\newblock


\bibitem[\protect\citeauthoryear{Hardy, Henecka, Ivey-Law, Nock, Patrini,
  Smith, and Thorne}{Hardy et~al\mbox{.}}{2017}]%
        {hardy2017private}
\bibfield{author}{\bibinfo{person}{Stephen Hardy}, \bibinfo{person}{Wilko
  Henecka}, \bibinfo{person}{Hamish Ivey-Law}, \bibinfo{person}{Richard Nock},
  \bibinfo{person}{Giorgio Patrini}, \bibinfo{person}{Guillaume Smith}, {and}
  \bibinfo{person}{Brian Thorne}.} \bibinfo{year}{2017}\natexlab{}.
\newblock \showarticletitle{Private federated learning on vertically
  partitioned data via entity resolution and additively homomorphic
  encryption}.
\newblock \bibinfo{journal}{\emph{arXiv preprint arXiv:1711.10677}}
  (\bibinfo{year}{2017}).
\newblock


\bibitem[\protect\citeauthoryear{He, Du, Wang, Tian, Tang, and Chua}{He
  et~al\mbox{.}}{2018}]%
        {he2018outer}
\bibfield{author}{\bibinfo{person}{Xiangnan He}, \bibinfo{person}{Xiaoyu Du},
  \bibinfo{person}{Xiang Wang}, \bibinfo{person}{Feng Tian},
  \bibinfo{person}{Jinhui Tang}, {and} \bibinfo{person}{Tat-Seng Chua}.}
  \bibinfo{year}{2018}\natexlab{}.
\newblock \showarticletitle{Outer product-based neural collaborative
  filtering}. In \bibinfo{booktitle}{\emph{IJCAI}}.
  \bibinfo{pages}{2227--2233}.
\newblock


\bibitem[\protect\citeauthoryear{Huang, He, Gao, Deng, Acero, and Heck}{Huang
  et~al\mbox{.}}{2013}]%
        {huang2013learning}
\bibfield{author}{\bibinfo{person}{Po-Sen Huang}, \bibinfo{person}{Xiaodong
  He}, \bibinfo{person}{Jianfeng Gao}, \bibinfo{person}{Li Deng},
  \bibinfo{person}{Alex Acero}, {and} \bibinfo{person}{Larry Heck}.}
  \bibinfo{year}{2013}\natexlab{}.
\newblock \showarticletitle{Learning deep structured semantic models for web
  search using clickthrough data}. In \bibinfo{booktitle}{\emph{CIKM}}. ACM,
  \bibinfo{pages}{2333--2338}.
\newblock


\bibitem[\protect\citeauthoryear{Jiang, Song, Tong, Wu, Zhao, Xu, and
  Yang}{Jiang et~al\mbox{.}}{2019}]%
        {jiang2019federated}
\bibfield{author}{\bibinfo{person}{Di Jiang}, \bibinfo{person}{Yuanfeng Song},
  \bibinfo{person}{Yongxin Tong}, \bibinfo{person}{Xueyang Wu},
  \bibinfo{person}{Weiwei Zhao}, \bibinfo{person}{Qian Xu}, {and}
  \bibinfo{person}{Qiang Yang}.} \bibinfo{year}{2019}\natexlab{}.
\newblock \showarticletitle{Federated Topic Modeling}. In
  \bibinfo{booktitle}{\emph{CIKM}}. \bibinfo{pages}{1071--1080}.
\newblock


\bibitem[\protect\citeauthoryear{Juan, Zhuang, Chin, and Lin}{Juan
  et~al\mbox{.}}{2016}]%
        {juan2016field}
\bibfield{author}{\bibinfo{person}{Yuchin Juan}, \bibinfo{person}{Yong Zhuang},
  \bibinfo{person}{Wei-Sheng Chin}, {and} \bibinfo{person}{Chih-Jen Lin}.}
  \bibinfo{year}{2016}\natexlab{}.
\newblock \showarticletitle{Field-aware factorization machines for CTR
  prediction}. In \bibinfo{booktitle}{\emph{RecSys.}} \bibinfo{pages}{43--50}.
\newblock


\bibitem[\protect\citeauthoryear{Li, Chen, Wang, Ren, Zhang, and Zhu}{Li
  et~al\mbox{.}}{2019}]%
        {li2019graph}
\bibfield{author}{\bibinfo{person}{Feng Li}, \bibinfo{person}{Zhenrui Chen},
  \bibinfo{person}{Pengjie Wang}, \bibinfo{person}{Yi Ren}, \bibinfo{person}{Di
  Zhang}, {and} \bibinfo{person}{Xiaoyu Zhu}.} \bibinfo{year}{2019}\natexlab{}.
\newblock \showarticletitle{Graph Intention Network for Click-through Rate
  Prediction in Sponsored Search}. In \bibinfo{booktitle}{\emph{SIGIR}}.
  \bibinfo{pages}{961--964}.
\newblock


\bibitem[\protect\citeauthoryear{Li, Cheng, Chen, Chen, and Wang}{Li
  et~al\mbox{.}}{2020}]%
        {li2020interpretable}
\bibfield{author}{\bibinfo{person}{Zeyu Li}, \bibinfo{person}{Wei Cheng},
  \bibinfo{person}{Yang Chen}, \bibinfo{person}{Haifeng Chen}, {and}
  \bibinfo{person}{Wei Wang}.} \bibinfo{year}{2020}\natexlab{}.
\newblock \showarticletitle{Interpretable Click-Through Rate Prediction through
  Hierarchical Attention}. In \bibinfo{booktitle}{\emph{WSDM}}.
  \bibinfo{pages}{313--321}.
\newblock


\bibitem[\protect\citeauthoryear{Lian, Zhou, Zhang, Chen, Xie, and Sun}{Lian
  et~al\mbox{.}}{2018}]%
        {lian2018xdeepfm}
\bibfield{author}{\bibinfo{person}{Jianxun Lian}, \bibinfo{person}{Xiaohuan
  Zhou}, \bibinfo{person}{Fuzheng Zhang}, \bibinfo{person}{Zhongxia Chen},
  \bibinfo{person}{Xing Xie}, {and} \bibinfo{person}{Guangzhong Sun}.}
  \bibinfo{year}{2018}\natexlab{}.
\newblock \showarticletitle{xdeepfm: Combining explicit and implicit feature
  interactions for recommender systems}. In \bibinfo{booktitle}{\emph{KDD}}.
  \bibinfo{pages}{1754--1763}.
\newblock


\bibitem[\protect\citeauthoryear{Liu, Kang, Zhang, Li, Cheng, Chen, Hong, and
  Yang}{Liu et~al\mbox{.}}{2019}]%
        {liu2019communication}
\bibfield{author}{\bibinfo{person}{Yang Liu}, \bibinfo{person}{Yan Kang},
  \bibinfo{person}{Xinwei Zhang}, \bibinfo{person}{Liping Li},
  \bibinfo{person}{Yong Cheng}, \bibinfo{person}{Tianjian Chen},
  \bibinfo{person}{Mingyi Hong}, {and} \bibinfo{person}{Qiang Yang}.}
  \bibinfo{year}{2019}\natexlab{}.
\newblock \showarticletitle{A Communication Efficient Vertical Federated
  Learning Framework}.
\newblock \bibinfo{journal}{\emph{arXiv preprint arXiv:1912.11187}}
  (\bibinfo{year}{2019}).
\newblock


\bibitem[\protect\citeauthoryear{Matteo and Dal~Zotto}{Matteo and
  Dal~Zotto}{2015}]%
        {matteo2015native}
\bibfield{author}{\bibinfo{person}{St{\'e}phane Matteo} {and}
  \bibinfo{person}{Cinzia Dal~Zotto}.} \bibinfo{year}{2015}\natexlab{}.
\newblock \showarticletitle{Native advertising, or how to stretch editorial to
  sponsored content within a transmedia branding era}.
\newblock In \bibinfo{booktitle}{\emph{Handbook of media branding}}.
  \bibinfo{publisher}{Springer}, \bibinfo{pages}{169--185}.
\newblock


\bibitem[\protect\citeauthoryear{McMahan, Moore, Ramage, Hampson, and
  y~Arcas}{McMahan et~al\mbox{.}}{2017}]%
        {mcmahan2017communication}
\bibfield{author}{\bibinfo{person}{Brendan McMahan}, \bibinfo{person}{Eider
  Moore}, \bibinfo{person}{Daniel Ramage}, \bibinfo{person}{Seth Hampson},
  {and} \bibinfo{person}{Blaise~Aguera y Arcas}.}
  \bibinfo{year}{2017}\natexlab{}.
\newblock \showarticletitle{Communication-Efficient Learning of Deep Networks
  from Decentralized Data}. In \bibinfo{booktitle}{\emph{AISTATS}}.
  \bibinfo{pages}{1273--1282}.
\newblock


\bibitem[\protect\citeauthoryear{Nock, Hardy, Henecka, Ivey-Law, Patrini,
  Smith, and Thorne}{Nock et~al\mbox{.}}{2018}]%
        {nock2018entity}
\bibfield{author}{\bibinfo{person}{Richard Nock}, \bibinfo{person}{Stephen
  Hardy}, \bibinfo{person}{Wilko Henecka}, \bibinfo{person}{Hamish Ivey-Law},
  \bibinfo{person}{Giorgio Patrini}, \bibinfo{person}{Guillaume Smith}, {and}
  \bibinfo{person}{Brian Thorne}.} \bibinfo{year}{2018}\natexlab{}.
\newblock \showarticletitle{Entity resolution and federated learning get a
  federated resolution}.
\newblock \bibinfo{journal}{\emph{arXiv preprint arXiv:1803.04035}}
  (\bibinfo{year}{2018}).
\newblock


\bibitem[\protect\citeauthoryear{Pan, Xu, Ruiz, Zhao, Pan, Sun, and Lu}{Pan
  et~al\mbox{.}}{2018}]%
        {pan2018field}
\bibfield{author}{\bibinfo{person}{Junwei Pan}, \bibinfo{person}{Jian Xu},
  \bibinfo{person}{Alfonso~Lobos Ruiz}, \bibinfo{person}{Wenliang Zhao},
  \bibinfo{person}{Shengjun Pan}, \bibinfo{person}{Yu Sun}, {and}
  \bibinfo{person}{Quan Lu}.} \bibinfo{year}{2018}\natexlab{}.
\newblock \showarticletitle{Field-weighted factorization machines for
  click-through rate prediction in display advertising}. In
  \bibinfo{booktitle}{\emph{WWW}}. \bibinfo{pages}{1349--1357}.
\newblock


\bibitem[\protect\citeauthoryear{Parsana, Poola, Wang, and Wang}{Parsana
  et~al\mbox{.}}{2018}]%
        {parsana2018improving}
\bibfield{author}{\bibinfo{person}{Mehul Parsana}, \bibinfo{person}{Krishna
  Poola}, \bibinfo{person}{Yajun Wang}, {and} \bibinfo{person}{Zhiguang Wang}.}
  \bibinfo{year}{2018}\natexlab{}.
\newblock \showarticletitle{Improving native ads ctr prediction by large scale
  event embedding and recurrent networks}.
\newblock \bibinfo{journal}{\emph{arXiv preprint arXiv:1804.09133}}
  (\bibinfo{year}{2018}).
\newblock


\bibitem[\protect\citeauthoryear{Pennington, Socher, and Manning}{Pennington
  et~al\mbox{.}}{2014}]%
        {pennington2014glove}
\bibfield{author}{\bibinfo{person}{Jeffrey Pennington},
  \bibinfo{person}{Richard Socher}, {and} \bibinfo{person}{Christopher
  Manning}.} \bibinfo{year}{2014}\natexlab{}.
\newblock \showarticletitle{Glove: Global vectors for word representation}. In
  \bibinfo{booktitle}{\emph{EMNLP}}. \bibinfo{pages}{1532--1543}.
\newblock


\bibitem[\protect\citeauthoryear{Pi, Bian, Zhou, Zhu, and Gai}{Pi
  et~al\mbox{.}}{2019}]%
        {pi2019practice}
\bibfield{author}{\bibinfo{person}{Qi Pi}, \bibinfo{person}{Weijie Bian},
  \bibinfo{person}{Guorui Zhou}, \bibinfo{person}{Xiaoqiang Zhu}, {and}
  \bibinfo{person}{Kun Gai}.} \bibinfo{year}{2019}\natexlab{}.
\newblock \showarticletitle{Practice on long sequential user behavior modeling
  for click-through rate prediction}. In \bibinfo{booktitle}{\emph{KDD}}.
  \bibinfo{pages}{2671--2679}.
\newblock


\bibitem[\protect\citeauthoryear{Qu, Cai, Ren, Zhang, Yu, Wen, and Wang}{Qu
  et~al\mbox{.}}{2016}]%
        {qu2016product}
\bibfield{author}{\bibinfo{person}{Yanru Qu}, \bibinfo{person}{Han Cai},
  \bibinfo{person}{Kan Ren}, \bibinfo{person}{Weinan Zhang},
  \bibinfo{person}{Yong Yu}, \bibinfo{person}{Ying Wen}, {and}
  \bibinfo{person}{Jun Wang}.} \bibinfo{year}{2016}\natexlab{}.
\newblock \showarticletitle{Product-based neural networks for user response
  prediction}. In \bibinfo{booktitle}{\emph{ICDM}}. IEEE,
  \bibinfo{pages}{1149--1154}.
\newblock


\bibitem[\protect\citeauthoryear{Ren, Yu, Yu, Yang, Yang, McCann, and
  Philip}{Ren et~al\mbox{.}}{2018}]%
        {ren2018textsf}
\bibfield{author}{\bibinfo{person}{Xuebin Ren}, \bibinfo{person}{Chia-Mu Yu},
  \bibinfo{person}{Weiren Yu}, \bibinfo{person}{Shusen Yang},
  \bibinfo{person}{Xinyu Yang}, \bibinfo{person}{Julie~A McCann}, {and}
  \bibinfo{person}{S~Yu Philip}.} \bibinfo{year}{2018}\natexlab{}.
\newblock \showarticletitle{High-Dimensional Crowdsourced Data Publication with
  Local Differential Privacy}.
\newblock \bibinfo{journal}{\emph{TIFS}} (\bibinfo{year}{2018}),
  \bibinfo{pages}{2151--2166}.
\newblock


\bibitem[\protect\citeauthoryear{Rendle}{Rendle}{2012}]%
        {rendle2012factorization}
\bibfield{author}{\bibinfo{person}{Steffen Rendle}.}
  \bibinfo{year}{2012}\natexlab{}.
\newblock \showarticletitle{Factorization machines with libfm}.
\newblock \bibinfo{journal}{\emph{TIST}} \bibinfo{volume}{3},
  \bibinfo{number}{3} (\bibinfo{year}{2012}), \bibinfo{pages}{57}.
\newblock


\bibitem[\protect\citeauthoryear{Rendle, Krichene, Zhang, and Anderson}{Rendle
  et~al\mbox{.}}{2020}]%
        {rendle2020neural}
\bibfield{author}{\bibinfo{person}{Steffen Rendle}, \bibinfo{person}{Walid
  Krichene}, \bibinfo{person}{Li Zhang}, {and} \bibinfo{person}{John
  Anderson}.} \bibinfo{year}{2020}\natexlab{}.
\newblock \showarticletitle{Neural Collaborative Filtering vs. Matrix
  Factorization Revisited}.
\newblock \bibinfo{journal}{\emph{arXiv preprint arXiv:2005.09683}}
  (\bibinfo{year}{2020}).
\newblock


\bibitem[\protect\citeauthoryear{Richardson, Dominowska, and Ragno}{Richardson
  et~al\mbox{.}}{2007}]%
        {richardson2007predicting}
\bibfield{author}{\bibinfo{person}{Matthew Richardson}, \bibinfo{person}{Ewa
  Dominowska}, {and} \bibinfo{person}{Robert Ragno}.}
  \bibinfo{year}{2007}\natexlab{}.
\newblock \showarticletitle{Predicting clicks: estimating the click-through
  rate for new ads}. In \bibinfo{booktitle}{\emph{WWW}}.
  \bibinfo{pages}{521--530}.
\newblock


\bibitem[\protect\citeauthoryear{Srivastava, Hinton, Krizhevsky, Sutskever, and
  Salakhutdinov}{Srivastava et~al\mbox{.}}{2014}]%
        {srivastava2014dropout}
\bibfield{author}{\bibinfo{person}{Nitish Srivastava},
  \bibinfo{person}{Geoffrey~E Hinton}, \bibinfo{person}{Alex Krizhevsky},
  \bibinfo{person}{Ilya Sutskever}, {and} \bibinfo{person}{Ruslan
  Salakhutdinov}.} \bibinfo{year}{2014}\natexlab{}.
\newblock \showarticletitle{Dropout: a simple way to prevent neural networks
  from overfitting.}
\newblock \bibinfo{journal}{\emph{JMLR}} \bibinfo{volume}{15},
  \bibinfo{number}{1} (\bibinfo{year}{2014}), \bibinfo{pages}{1929--1958}.
\newblock


\bibitem[\protect\citeauthoryear{Vaswani, Shazeer, Parmar, Uszkoreit, Jones,
  Gomez, Kaiser, and Polosukhin}{Vaswani et~al\mbox{.}}{2017}]%
        {vaswani2017attention}
\bibfield{author}{\bibinfo{person}{Ashish Vaswani}, \bibinfo{person}{Noam
  Shazeer}, \bibinfo{person}{Niki Parmar}, \bibinfo{person}{Jakob Uszkoreit},
  \bibinfo{person}{Llion Jones}, \bibinfo{person}{Aidan~N Gomez},
  \bibinfo{person}{{\L}ukasz Kaiser}, {and} \bibinfo{person}{Illia
  Polosukhin}.} \bibinfo{year}{2017}\natexlab{}.
\newblock \showarticletitle{Attention is all you need}. In
  \bibinfo{booktitle}{\emph{NIPS}}. \bibinfo{pages}{5998--6008}.
\newblock


\bibitem[\protect\citeauthoryear{Wojdynski and Evans}{Wojdynski and
  Evans}{2016}]%
        {wojdynski2016going}
\bibfield{author}{\bibinfo{person}{Bartosz~W Wojdynski} {and}
  \bibinfo{person}{Nathaniel~J Evans}.} \bibinfo{year}{2016}\natexlab{}.
\newblock \showarticletitle{Going native: Effects of disclosure position and
  language on the recognition and evaluation of online native advertising}.
\newblock \bibinfo{journal}{\emph{Journal of Advertising}}
  \bibinfo{volume}{45}, \bibinfo{number}{2} (\bibinfo{year}{2016}),
  \bibinfo{pages}{157--168}.
\newblock


\bibitem[\protect\citeauthoryear{Wu, Wu, Ge, Qi, Huang, and Xie}{Wu
  et~al\mbox{.}}{2019}]%
        {wu2019nrms}
\bibfield{author}{\bibinfo{person}{Chuhan Wu}, \bibinfo{person}{Fangzhao Wu},
  \bibinfo{person}{Suyu Ge}, \bibinfo{person}{Tao Qi},
  \bibinfo{person}{Yongfeng Huang}, {and} \bibinfo{person}{Xing Xie}.}
  \bibinfo{year}{2019}\natexlab{}.
\newblock \showarticletitle{Neural News Recommendation with Multi-Head
  Self-Attention}. In \bibinfo{booktitle}{\emph{EMNLP}}.
  \bibinfo{pages}{6390--6395}.
\newblock


\bibitem[\protect\citeauthoryear{Yang, Liu, Chen, and Tong}{Yang
  et~al\mbox{.}}{2019}]%
        {yang2019federated}
\bibfield{author}{\bibinfo{person}{Qiang Yang}, \bibinfo{person}{Yang Liu},
  \bibinfo{person}{Tianjian Chen}, {and} \bibinfo{person}{Yongxin Tong}.}
  \bibinfo{year}{2019}\natexlab{}.
\newblock \showarticletitle{Federated machine learning: Concept and
  applications}.
\newblock \bibinfo{journal}{\emph{TIST}} \bibinfo{volume}{10},
  \bibinfo{number}{2} (\bibinfo{year}{2019}), \bibinfo{pages}{1--19}.
\newblock


\bibitem[\protect\citeauthoryear{Yang, Yang, Dyer, He, Smola, and Hovy}{Yang
  et~al\mbox{.}}{2016}]%
        {yang2016hierarchical}
\bibfield{author}{\bibinfo{person}{Zichao Yang}, \bibinfo{person}{Diyi Yang},
  \bibinfo{person}{Chris Dyer}, \bibinfo{person}{Xiaodong He},
  \bibinfo{person}{Alex Smola}, {and} \bibinfo{person}{Eduard Hovy}.}
  \bibinfo{year}{2016}\natexlab{}.
\newblock \showarticletitle{Hierarchical attention networks for document
  classification}. In \bibinfo{booktitle}{\emph{NAACL-HLT}}.
  \bibinfo{pages}{1480--1489}.
\newblock


\bibitem[\protect\citeauthoryear{Zhou, Mou, Fan, Pi, Bian, Zhou, Zhu, and
  Gai}{Zhou et~al\mbox{.}}{2019}]%
        {zhou2019deep}
\bibfield{author}{\bibinfo{person}{Guorui Zhou}, \bibinfo{person}{Na Mou},
  \bibinfo{person}{Ying Fan}, \bibinfo{person}{Qi Pi}, \bibinfo{person}{Weijie
  Bian}, \bibinfo{person}{Chang Zhou}, \bibinfo{person}{Xiaoqiang Zhu}, {and}
  \bibinfo{person}{Kun Gai}.} \bibinfo{year}{2019}\natexlab{}.
\newblock \showarticletitle{Deep interest evolution network for click-through
  rate prediction}. In \bibinfo{booktitle}{\emph{AAAI}},
  Vol.~\bibinfo{volume}{33}. \bibinfo{pages}{5941--5948}.
\newblock


\bibitem[\protect\citeauthoryear{Zhou, Zhu, Song, Fan, Zhu, Ma, Yan, Jin, Li,
  and Gai}{Zhou et~al\mbox{.}}{2018}]%
        {zhou2018deep}
\bibfield{author}{\bibinfo{person}{Guorui Zhou}, \bibinfo{person}{Xiaoqiang
  Zhu}, \bibinfo{person}{Chenru Song}, \bibinfo{person}{Ying Fan},
  \bibinfo{person}{Han Zhu}, \bibinfo{person}{Xiao Ma},
  \bibinfo{person}{Yanghui Yan}, \bibinfo{person}{Junqi Jin},
  \bibinfo{person}{Han Li}, {and} \bibinfo{person}{Kun Gai}.}
  \bibinfo{year}{2018}\natexlab{}.
\newblock \showarticletitle{Deep interest network for click-through rate
  prediction}. In \bibinfo{booktitle}{\emph{KDD}}. \bibinfo{pages}{1059--1068}.
\newblock


\end{thebibliography}

%\clearpage
%\input{data/supplement}
\end{document}